\documentclass[twocolumn,showpacs,preprintnumbers,amsmath,amssymb,prl,superscriptaddress,longbibliography]{revtex4-1}
\usepackage{bbm}
\usepackage{mathrsfs}
\usepackage{graphicx}
\usepackage{dcolumn}
\usepackage{bm}
\usepackage{amsmath}
\usepackage{amsfonts}
\usepackage{color}
\usepackage[colorlinks=true,linkcolor=magenta,urlcolor=magenta,citecolor=cyan,anchorcolor=blue]{hyperref}
\usepackage{floatrow}

\begin{document}

\title{Monolayer V$_2$\textit{MX}$_4$: A New Family of Quantum Anomalous Hall Insulators}
\author{Yadong Jiang}
\affiliation{State Key Laboratory of Surface Physics and Department of Physics, Fudan University, Shanghai 200433, China}
\author{Huan Wang}
\affiliation{State Key Laboratory of Surface Physics and Department of Physics, Fudan University, Shanghai 200433, China}
\author{Kejie Bao}
\affiliation{State Key Laboratory of Surface Physics and Department of Physics, Fudan University, Shanghai 200433, China}
\author{Zhaochen Liu}
\affiliation{State Key Laboratory of Surface Physics and Department of Physics, Fudan University, Shanghai 200433, China}
\author{Jing Wang}
\thanks{Corresponding author: wjingphys@fudan.edu.cn}
\affiliation{State Key Laboratory of Surface Physics and Department of Physics, Fudan University, Shanghai 200433, China}
\affiliation{Institute for Nanoelectronic Devices and Quantum Computing, Zhangjiang Fudan International Innovation Center, Fudan University, Shanghai 200433, China}
\affiliation{Hefei National Laboratory, Hefei 230088, China}

\begin{abstract}
We theoretically propose that the van der Waals layered ternary transition metal chalcogenide V$_2$\textit{MX}$_4$ ($M=$ W, Mo; $X=$ S, Se) is a new family of quantum anomalous Hall insulators with sizable bulk gap and Chern number $\mathcal{C}=-1$. The large topological gap originates from the \emph{deep} band inversion between spin up bands contributed by $d_{xz},d_{yz}$ orbitals of V and spin down band from $d_{z^2}$ orbital of $M$ at Fermi level. Remarkably, the Curie temperature of monolayer V$_2$\textit{MX}$_4$ is predicted to be much higher than that of monolayer MnBi$_2$Te$_4$. Furthermore, the thickness dependence of the Chern number for few multilayers shows interesting oscillating behavior. The general physics from the $d$-orbitals here applies to a large class of ternary transition metal chalcogenide such as Ti$_2$W$X_4$ with the space group $P$-$42m$. These interesting predictions, if realized experimentally, could greatly promote the research and application of topological quantum physics.
\end{abstract} 

\date{\today}

%\pacs{
%        73.22.-f  % Electronic structure of nanoscale materials and related systems
%        02.20.-a  % Group theory
%        73.43.-f  % Quantum Hall effects
%      }

\maketitle

{\color{blue}\emph{Introduction.}}
The discovery of the quantum anomalous Hall (QAH) effect set a remarkable example for understanding topological states of quantum matter in condensed matter physics and material science~\cite{hasan2010,qi2011,tokura2019,wang2017c,bernevig2022,chang2023}. Such a state is characterized by a topologically nontrivial insulating bulk with a finite Chern number $\mathcal{C}$~\cite{thouless1982,haldane1988} but gapless chiral edge states, which is promising for the realization of dissipationless electronic devices~\cite{zhang2012,wang2013a} and topological computation~\cite{qi2010b,wang2015c,lian2018b}. The QAH effect has been observed first in magnetically doped topological insulators (TI)~\cite{chang2013b,chang2015,mogi2015,bestwick2015,watanabe2019}, and subsequently in the intrinsic magnetic TI MnBi$_2$Te$_4$~\cite{deng2020}, in the moir\'e graphene~\cite{serlin2020} and moir\'e transition metal dichalcogenide~\cite{li2021}, but only at low temperature (below $5$~K). Such low critical temperature is a weighty obstacle for practical applications, for example the quantum resistance standard~\cite{okazaki2022}. Seeking new QAH insulator materials~\cite{you2019,sunj2020,liy2020,xuan2022,sun2020,li2022,jiang2023} with preferably large bulk gaps has become an important goal in topological material research.

Physically, the basic mechanism for the QAH effect is band inversion of the spin polarized bands~\cite{liu2008}, where both the spin-orbit coupling (SOC) and ferromagnetism are sufficiently strong. From the materials perspective, strong SOC prefers heavy elements, while the ferromagnetism favors transition metal elements preferably with 3$d$ electrons. Thus the challenge in searching for large-gap QAH insulator materials is to synergize the seemingly conflicting requirement of SOC and ferromagnetism. Indeed, the inhomogeneities in magnetic TI~\cite{yu2010,wang2015d,zhang2019,li2019,otrokov2019a} from magnetic dopants~\cite{chong2020} and defects~\cite{garnica2022} drastically suppress the exchange gap by several order of magnitude, which fundamentally limits the exactly quantized anomalous Hall effect to very low temperatures. Therefore, finding stoichiometric 2D magnetic materials for QAH effect preferably in monolayer with versatile tunability is highly desired.
 
\begin{figure}[b]
\begin{center}
\includegraphics[width=3.4in, clip=true]{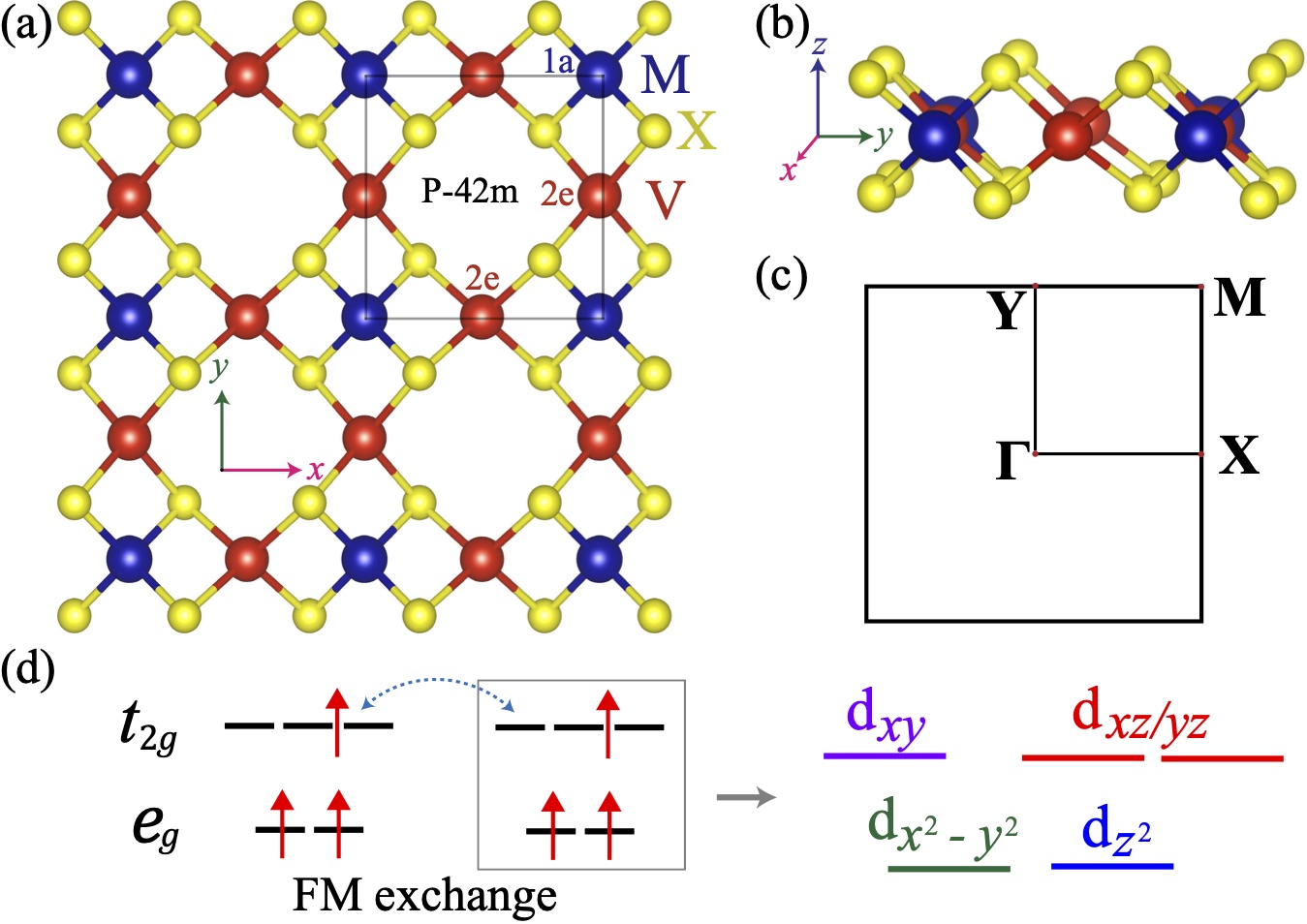}
\end{center}
\caption{(a),(b) Atomic structure of monolayer V$_2$\textit{MX}$_4$ from top and side views. The Wyckoff positions $1a$ and $2e$ are displayed (notation adopted from Bilbao Crystallographic Server~\cite{bilbao2,bilbao3,bilbao1,slager2017,vergniory2017,elcoro2017,bradlyn2017}). The magnetic ground state of the 2D materials class is FM along $z$ direction with spin magnetic moment $2.6\mu_B$ per V atom. The key symmetry operations of $P$-$42m$ include $S_{4z}$, $C_{2x}$ and $C_{2z}$ rotations, where $S_{4z}\equiv\mathcal{I}C_{4z}$ and  $\mathcal{I}$ is inversion symmetry. (c) Brillouin zone. (d) Crystal field splitting and schematic diagram of the FM kinetic exchange coupling between the V atoms.}
\label{fig1}
\end{figure} 
 
Here we predict a series of large-gap QAH insulators in monolayer V$_2$\textit{MX}$_4$ ($M=$ W, Mo; $X=$ S, Se), based on density functional theory (DFT) calculations and tight-binding model. The Vienna \emph{ab initio} simulation package~\cite{kresse1996} is employed by using the Perdew-Burke-Ernzerhof~\cite{perdew1996} generalized-gradient approximation. We perform $\text{DFT}+\text{Hubbard}$ $U$ calculations~\cite{dudarev1998}. The predicted topology was further verified by Heyd-Scuseria-Ernzerhof (HSE) hybrid functional with band gap listed in Table~\ref{tab1}~\cite{krukau2006}. These materials have the ferromagnetic (FM) ground state with Chern number $\mathcal{C}=-1$ and extraordinarily large bulk gaps ($\sim0.2$~eV). We find the large topological gap originates from the deep band inversion between V $d_{xz},d_{yz}$ orbitals and $M$ $d_{z^2}$ orbital. The rich choice of candidate materials in Table~\ref{tab1} indicates that the physics here is generic with the space group $P$-$42m$.

{\color{blue}\emph{Structure and magnetic properties.}} The monolayer V$_2$\textit{MX}$_4$ has a tetragonal lattice with the space group $P$-$42m$ (No.~111). As shown in Fig.~\ref{fig1}(a), each primitive cell includes three (i.e., one V$_2M$ and two $X_2$) atomic layers, where each V or $M$ atom is surrounded by four $X$ atoms forming a distorted edge-sharing tetrahedron. These QAH materials are obtained from high-throughput screening of insulating V$_2$\textit{MX}$_4$ with $M$ from group 5 and 6,  and $X$ from group 16. Their lattice constants are listed in Table~\ref{tab1}. The dynamical and thermal stability of monolayer V$_2$\textit{MX}$_4$ are confirmed by first-principles phonon and molecular dynamics calculations~\cite{supple}, respectively. We will mainly discuss V$_2$WS$_4$ with similar results for other materials in this class. In reality, Cu$_2\textit{MX}_4$ and Ag$_2\textit{MX}_4$ with the same structure have been experimentally synthesized~\cite{crossland2005,gan2014,hu2016,zhan2018,wu2019,lin2019}, which implies high probability to fabricate V$_2$\textit{MX}$_4$. Meanwhile, the van der Waals nature of these materials implies the experimental feasibility to exfoliate monolayer from bulk sample.

\begin{table}[t]
\caption{Lattice constant; Magnetocrystalline anisotropy energy (MAE) per unit cell $E_{\text{MAE}}$, defined as the total energy difference between in-plane and out-of-plane spin configurations; Curie temperature $T_c$ from Monte Carlo simulations; Band gap $E_g$ by using HSE method.}
\begin{center}\label{tab1}
\renewcommand{\arraystretch}{1.3}
\begin{tabular*}{3.4in}
{@{\extracolsep{\fill}}ccccc}
\hline
\hline
Materials&$a$ (\r{A})& $T_c$ (K)& $E_{\rm{MAE}}$ (meV)& $E_g$ (meV)\\
\hline
V$_2$WS$_4$  & 5.74 & 470 & 12.1 & 279\\
V$_2$WSe$_4$ & 5.82 & 440 & 13.2 & 258\\
V$_2$MoS$_4$ & 5.72 & 310 & 2.0  & 115\\
V$_2$MoSe$_4$& 5.83 & 284 & 2.2  & 70\\
Ti$_2$WS$_4$ & 5.75 & 240 & 10.7 & 259\\
Ti$_2$WSe$_4$& 5.79 & 210 & 13.7 & 275\\
\hline
\hline
\end{tabular*}
\end{center}
\end{table}

\begin{figure}[t]
\begin{center}
\includegraphics[width=3.4in, clip=true]{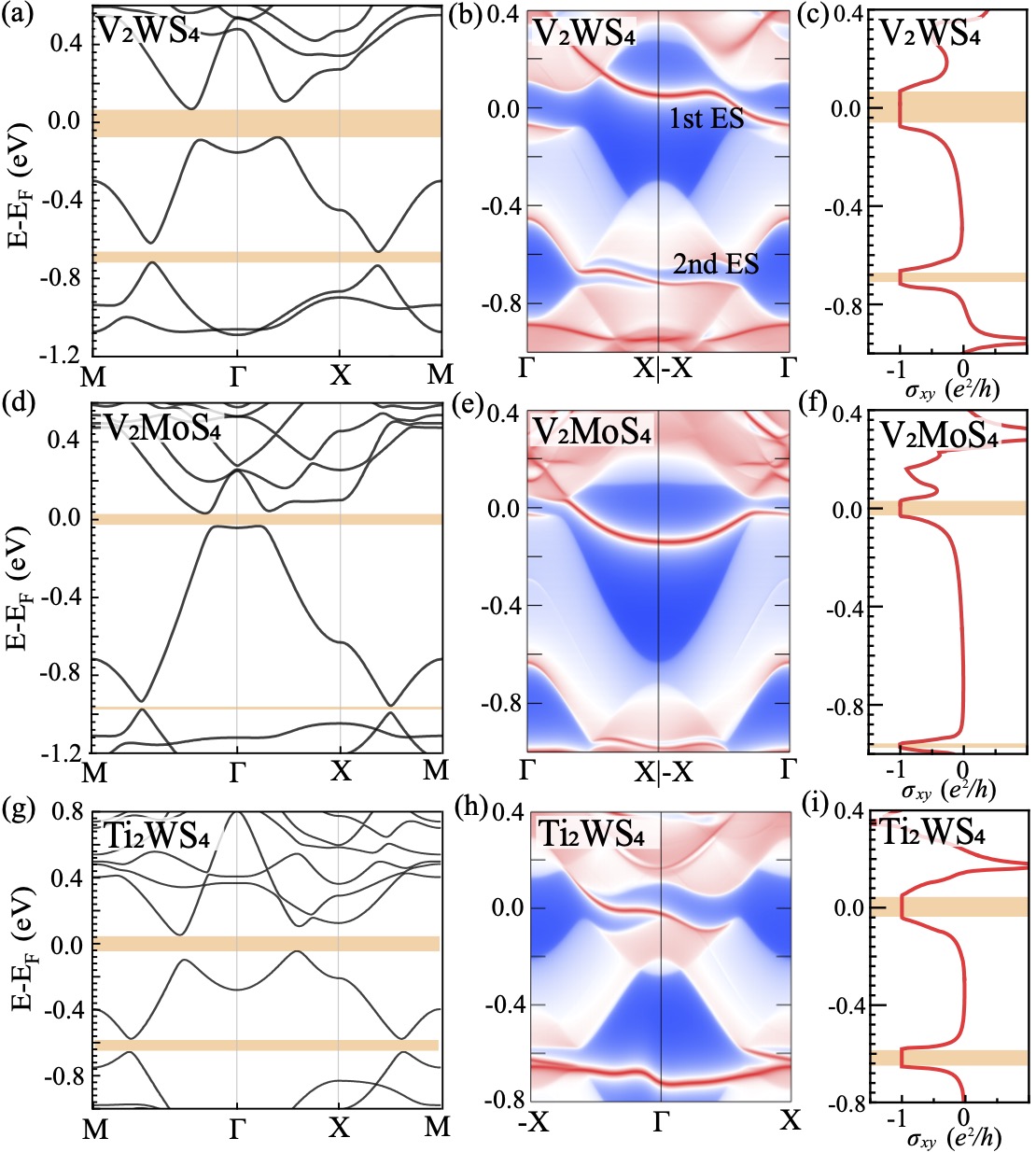}
\end{center}
\caption{Electronic structures and topological properties of monolayer V$_2$WS$_4$, V$_2$MoS$_4$ and Ti$_2$WS$_4$. (a)-(c) V$_2$WS$_4$, (d)-(f) V$_2$MoS$_4$, (g)-(i) Ti$_2$WS$_4$, The band structure with SOC; topological edge states (ES) calculated along $x$ axis; and anomalous Hall conductance $\sigma_{xy}$ as a function of Fermi energy, respectively. The shaded regions in (a),(d),(g) denote the topological gap.}
\label{fig2}
\end{figure}

\begin{figure}[t]
\begin{center}
\includegraphics[width=3.4in, clip=true]{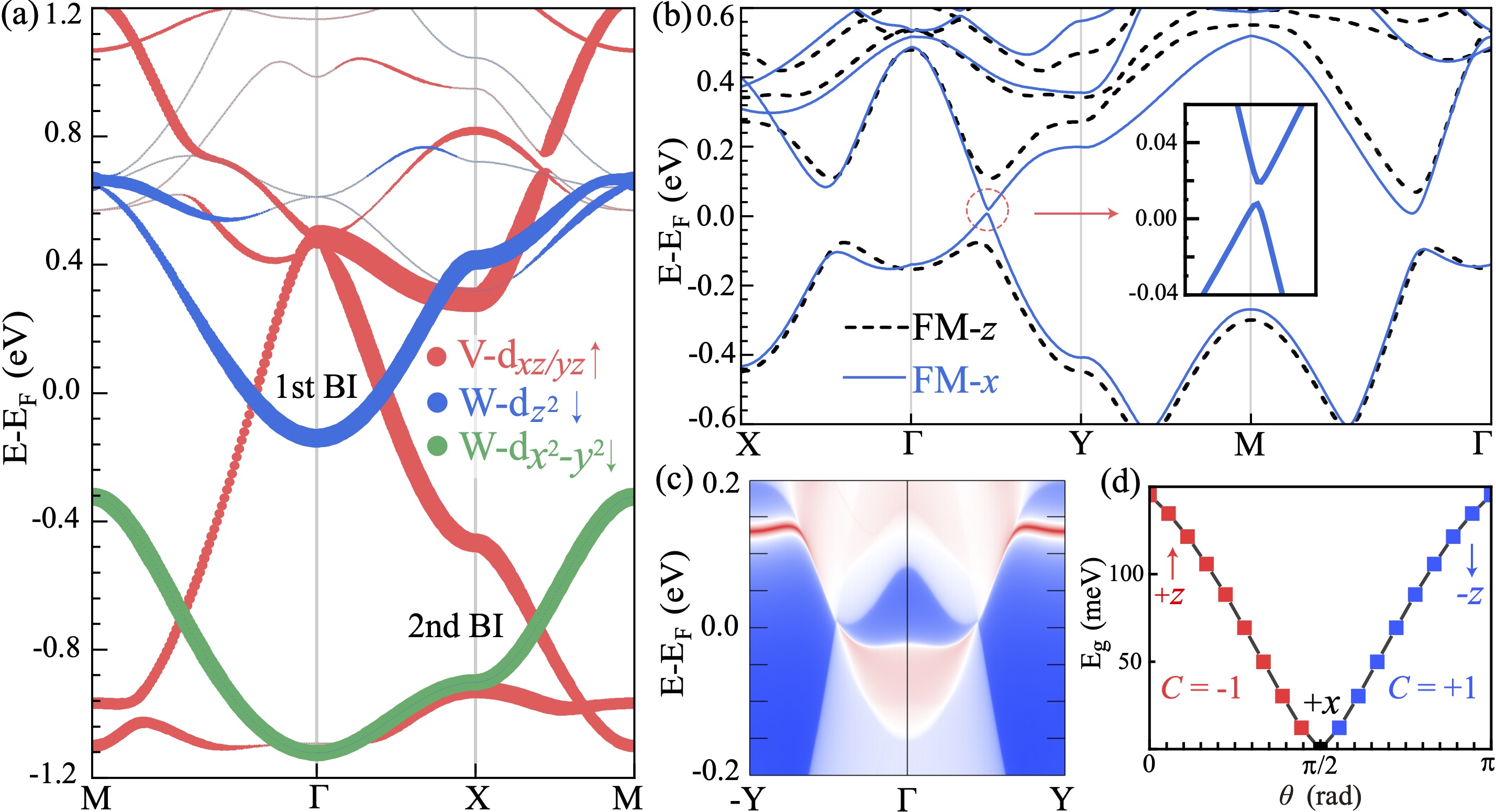}
\end{center}
\caption{(a) The $d$-orbitals projection band structures without SOC of monolayer V$_2$WS$_4$, only those bands which is related to band inversion are shown. (b) The band structure for FM along $z$- and $x$-axis of V$_2$WS$_4$ with SOC. (c) Energy and momentum dependence of the edge local density of state along $y$-axis under FM-$x$ state of V$_2$WS$_4$. (d) Dependence of bulk gap along the high symmetry line and $\mathcal{C}$ on the spin orientation quantified by a polar angle $\theta$, where $\theta=0, \pi/2, \pi$ denote the $+z, +x, -z$ directions, respectively. }
\label{fig3}
\end{figure}

First-principles calculations show V$_2MX_4$ listed in Table~\ref{tab1} have strong FM ground state with an out-of-plane easy axis~\cite{supple}. The underlying mechanism of FM can be elucidated from orbital occupation. The magnetic moments are mainly provided by V ($\approx2.6\mu_B$) rather than W ($\approx0.4\mu_B$). The fractional magnetic moment is due to band inversion between V $d_{xz,yz}$ orbitals and W $d_{z^2}$ orbital (Fig.~\ref{fig3}(a)). Thus the magnetism is from V atoms. The tetrahedral crystal field splits V $3d$ orbitals into doublet $e_g (d_{x^2-y^2},d_{z^2})$ and triplet $t_{2g}(d_{xy},d_{xz},d_{yz})$ orbitals (Fig.~\ref{fig1}(d)). The energy of $e_g$ stays lower with respect to $t_{2g}$, because the latter point towards the negatively charged ligands. Thus each V atom is in the $e^2_gt_{2g}^1$ configuration with the magnetic moment of $3\mu_B$ according to the Hund's rule, which is close to the DFT calculations. The FM exchange coupling between neighboring V atoms is strongly enhanced by Hund's rule interaction due to empty $t_{2g}$ orbitals~\cite{khomskii2004}. Furthermore, the predicted Curie temperature for monolayer V$_2$\textit{MX}$_4$ is much higher than that of MnBi$_2$Te$_4$.

{\color{blue}\emph{Electronic structures.}} 
Fig.~\ref{fig2}(a) and Fig.~\ref{fig3}(a) display the electronic structure of monolayer V$_2$WS$_4$ with and without SOC, respectively. There are two band inversions between different spin polarized bands, both of which are further gapped by SOC. Specifically, one is near the Fermi energy $E_F$ between spin up bands contributed by $d_{xy},d_{yz}$ orbitals of V and spin down band by $d_{z^2}$ orbital of W, the other is about $0.7$~eV below $E_F$ between V $d_{xy},d_{yz}$ spin up bands and W $d_{x^2-y^2}$ spin down band. There also exists a spin polarized quadratic band touching at $\Gamma$ point from $d_{xy},d_{yz}$ orbitals of V above $E_F$, with the nontrivial gap opened by SOC~\cite{wang2021}. The anomalous Hall conductance $\sigma_{xy}$ versus $E_F$ is calculated in Fig.~\ref{fig2}(c), which displays a quantized value of $-e^2/h$ near both $E_F$ and $E_F-0.7$~eV. This indicates the topological nontrivial bands with $\mathcal{C}=-1$ below $E_F-0.7$~eV, which is consistent with single chiral edge states dispersing within the bulk gap as in the edge local density of states (Fig.~\ref{fig2}(b)). Interestingly, there also exists an occupied 2nd chiral edge state which is $0.7$~eV below $E_F$, which can be measured by scanning tunneling microscope. By replacing W by the same group element Mo, V$_2$MoS$_4$ monolayer has similar band structure and same topological properties as shown in Fig.~\ref{fig2}(d)-\ref{fig2}(f). The topological gap of monolayer V$_2$W$X_4$ is larger than that of V$_2$Mo$X_4$, which is due to enhanced SOC from heavier elements and deeper band inversion.

The topological gap strongly depends on spin orientations. Fig.~\ref{fig3}(b) shows the band structure for the in-plane ferromagnetism along $x$-axis. The symmetries of the system reduces to $C_{2x}$ and $C_{2z}\mathcal{T}$, where $\mathcal{T}$ is time-reversal. One can see the gap along $\Gamma$-Y decreases but not closes, for there is no symmetry to guarantee a gapless point. Along the high symmetry lines, there is a small negative indirect gap between valence top along $\Gamma$-Y and conduction bottom along $\Gamma$-M. The Chern number of valence bands are calculated to be $\mathcal{C}=0$, which is guaranteed by $C_{2z}\mathcal{T}$. Similar to $\mathcal{IT}$, $\sigma_{xy}$ is odd under $C_{2z}\mathcal{T}$ in 2D materials, which ensures $\mathcal{C}=0$ when ferromagnetism is completely along any in-plane direction. This is consistent with the edge local density of state calculation in Fig.~\ref{fig3}(c), where there is no chiral edge state. The FM-$x$ state is trivial without any kinds of topology, which is checked by the band representation of $C_{2x}$ (i.e., symmetry indicator). Then by varying the spin orientation from $+z$ to $+x$, then to $-z$ axis, the band gap monotonically decreases to close, and negative then reopens, which is accompanied by the topological phase transitions from $\mathcal{C}=-1$ to $\mathcal{C}=0$ and then to $\mathcal{C}=1$ in Fig.~\ref{fig3}(d). The gap varies approximately in relation as $E_g\propto S\cos(\theta+\theta_0)\approx \langle S_z\rangle$, with $\theta_0\approx 0.1$~\cite{supple}. This simply means the topologically nontrivial gap is opened by SOC related to $\langle S_z\rangle$ component, while the trivial gap is associated with $\langle S_x\rangle$ component.

The intimate relationship between band gap and spin orientation implies that the electronic structure is renormalized by magnon excitations. In FM-$z$ ground state, the magnons contribution to magnetization is $\langle S_z\rangle=S-\int d^2\mathbf{k}n_B(\epsilon_\mathbf{k})/(2\pi)^2$, where $n_B(\epsilon_{\mathbf{k}})\equiv1/[\exp(\epsilon_{\mathbf{k}}/k_BT)-1]$ is the Bose distribution, $\epsilon_{\mathbf{k}}$ is magnon dispersion. Taking V$_2$WS$_4$ as an example, $S=3/2$ and the magnon gap is estimated to be $5.34$~meV. Thus when $T<58$~K, the magnon is absent; when $T\lesssim200$~K, the reduction in $\langle S_z\rangle$ from magnon excitation is less than $6\%$ compared to $\langle S_z\rangle=3/2$~\cite{supple}. Therefore, the large topological gap of ground state still holds in the presence of magnon excitation. It is worth mentioning that when temperature is close to $T_c$, the significant thermal spin fluctuation will decrease $\langle S_z\rangle$ and topological gap dramatically.

{\color{blue}\emph{Tight-binding model and origin of topology.}} 
To reveal the origin of $\mathcal{C}=-1$ topology in the electronic structure, we perform the symmetry analysis of the band irreducible representation and construct a tight-binding model to recover the essential topological physics. The first and second band inversions occur at $\Gamma$ and M point, respectively. Naively, one may simply count the angular momentum difference locally at the band inversion point to be the Chern number change as in the conventional $s$-$p$ band inversion. However, this is incorrect due to band inversion from the opposite spin polarization here. For example, at M point for second band inversion, only spin contributes to the angular momentum difference, which implies the lowest band ($d^\downarrow_{x^2-y^2}$ from W) should have $\mathcal{C}=1$, this is contrary to the first principles calculations of $\mathcal{C}=-1$.

Then we construct a concrete tight-binding model including $d^\uparrow_{xz},d^\uparrow_{yz}$ of V and $d_{z^2}^\downarrow, d_{x^2-y^2}^\downarrow$ of W to decipher the origin of topology. There are two V in an unit cell, and $d_{xz},d_{yz}$ orbitals of each V are non-degenerate. However, $d_{xz}$ of one V and $d_{yz}$ of the other V are degenerate, which are related to each other by $S_{4z}$. Therefore, for the low energy physics, we only need to consider $d^\uparrow_{xz},d^\uparrow_{yz}$ from two V, respectively and $d_{z^2}^\downarrow, d_{x^2-y^2}^\downarrow$ of W, namely a four orbitals model. The Hamiltonian is obtained by considering the nearest-neighbor and next-nearest-neighbor hopping with SOC included, where the explicit forms are in Supplementary Materials~\cite{supple}. As shown in Fig.~\ref{fig4}(a) and~\ref{fig4}(b), the band structure and the corresponding irreducible representations of high symmetry points (listed in Table~\ref{tab2}) in DFT calculation (Fig.~\ref{fig2}(a) and Fig.~\ref{fig3}(a)) are rebuilt in our tight-binding model.

\begin{figure}[t]
\begin{center}
\includegraphics[width=3.3in, clip=true]{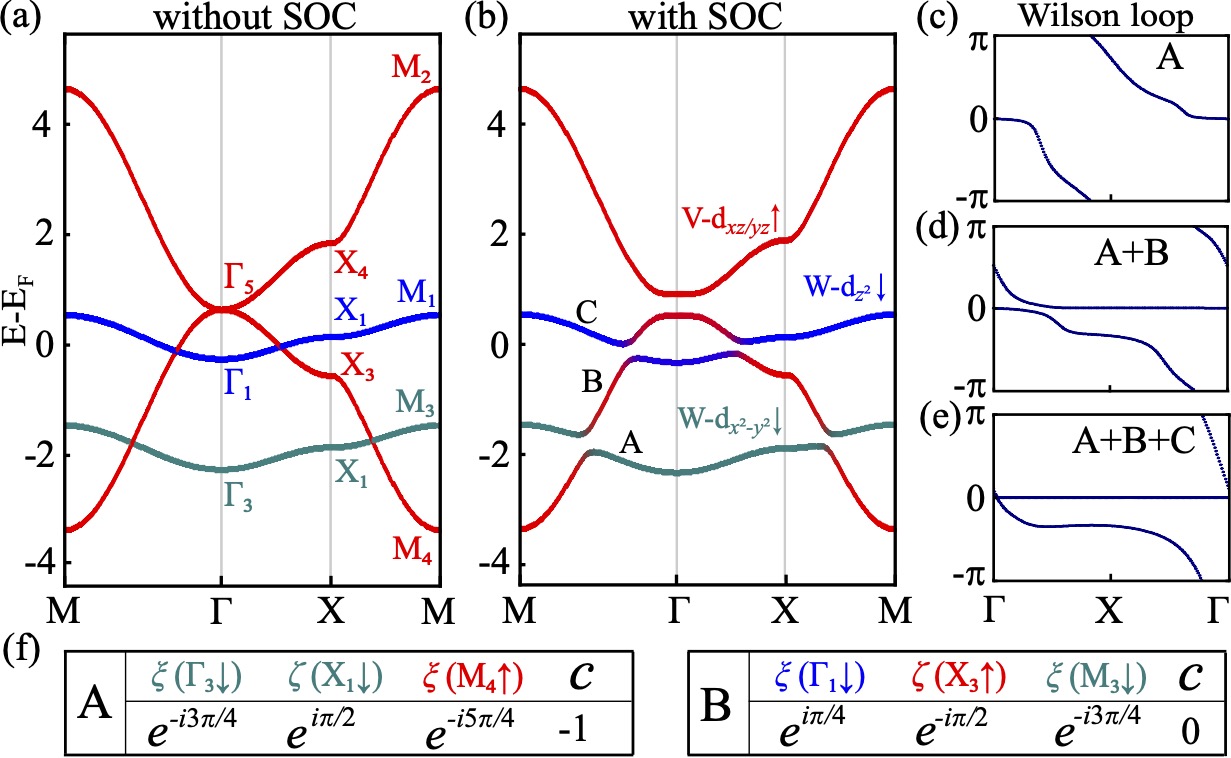}
\end{center}
\caption{(a),(b) The $d$-orbitals projection band structure of the tight-binding model with and without SOC. The irreducible representation at high-symmetry points of the Brillouin zone boundary and the orbital compositions (color of the bands) are consistent with that in Fig.~\ref{fig3}(a). (c)-(e), The Wilson loop for the lowest one, two and three bands in (b), respectively. (f) Eigenvalues of $S_{4z}$ at $\Gamma$ and M points and $C_{2z}$ at X point of bands A and B. Chern number $\mathcal{C}$ are calculated by Eq.~(\ref{eq1}).}
\label{fig4}
\end{figure}

The symmetry generators of space group $P$-$42m$ are $S_{4z}$, $C_{2z}$ and $C_{2x}$. For tight-binding model of V$_2$\textit{MX}$_4$, $\mathcal{I}$ can be viewed as $C_{2z}$ due to lacking of $z$ direction. Then $S_{4z}$ becomes $C^3_{4z}$, namely, the system is effectively $C_{4z}$ invariant. The Chern number of band in a $S_{4}$ invariant system is shown to be~\cite{fang2012,supple},
\begin{equation}\label{eq1}
i^\mathcal{C}=\prod \limits_{j\in\text{occupied}} (-1)^F \xi_j(\Gamma)\xi_j(M)\zeta_j(X).
\end{equation}
Here $F=1$ for spinful case. $\xi_j(k)$ is the eigenvalue of $S_{4z}$ at the $S_{4z}$-invariant $\Gamma$ and M points of the $j$-th band, $\zeta_j(k)$ is the eigenvalue of $C_{2z}$ at the $C_{2z}$-invariant X point on the $j$-th band. Now the band topology can be diagnosed by the symmetry information listed in Table~\ref{tab2}. The eigenvalues and Chern numbers for lowest (A) and second lowest band (B) are calculated in Fig.~\ref{fig4}(f). All of these are consistent with the Wilson loop calculations for the lowest one and two bands shown in Fig.~\ref{fig4}(c) and~\ref{fig4}(d). The quadratic band touching at $\Gamma_5$ is from degenerate $d_{xz},d_{yz}$, which have an effective angular momentum $\ell_z=\pm1$, and SOC opens a topologically nontrivial gap~\cite{wang2021}. Thus the Chern number summation of lowest three bands is only determined by the sign of SOC. As shown in Fig.~\ref{fig4}(e), the total Chern number of the lowest three bands is $\mathcal{C}=-1$, namely, both bands B and C have $\mathcal{C}=0$, and the only nontrivial $\mathcal{C}=-1$ is carried by band A, which is consistent with first principles calculations. If we artificially reverse the sign of SOC, then the Chern number of the lowest three bands are $(\mathcal{C}_{A},\mathcal{C}_B,\mathcal{C}_C)=(-1,0,2)$. Now we fully understand that the topology in this system is not only from gapping the degenerate $d_{xz},d_{yz}$ orbitals of V by SOC, but also the band inversions from them and W $d_{z^2}$ and $d_{x^2-y^2}$.

Here we analyze the origin of the large topological gap. The topological gap is from SOC term as $\lambda_{\text{so}}\boldsymbol{\ell}\cdot\mathbf{s}=\lambda_{\text{so}}(\ell_+s_-+\ell_-s_+)/2+\lambda_{\text{so}}\ell_zs_z$, where $\lambda_{\text{so}}$ is atomic SOC strength. The nontrivial gap is opened by combining the first term with orbital hopping and hybridization. For spin polarized band inversion in QAH, two bands are inverted at certain high-symmetry point in the Brillouin zone, and topological gap $E_g$ opens at a finite wave vector $\delta k$ away from the band inversion point due to orbital hybridization with $E_g\propto\lambda'\delta k$, where $\lambda'$ is the hybridization strength. For spin polarized band inversion with same spin such as MnBi$_2$Te$_4$~\cite{zhang2019,li2019}, $\lambda'\propto\lambda_{\text{so}}^2$, namely second-order process. In V$_2$\textit{MX}$_4$, the band inversion are between opposite spin polarized bands, $\lambda'\propto\lambda_{\text{so}}$, thus a lower-order process with greater strength in gap opening. Meanwhile, these materials have deep band inversion with large $\delta k$. Therefore, both $\lambda'$ and $\delta k$ are enhanced in V$_2$\textit{MX}$_4$ and lead to large topological gap.

\begin{table}[t]
\caption{Partial elementary band representations without time-reversal symmetry for space group $P$-$42m$.} 
\begin{center}\label{tab2}
\renewcommand{\arraystretch}{1.4}
\begin{tabular*}{3.4in}
{@{\extracolsep{\fill}}c|ccc}
\hline
\hline
  & $\Gamma$ & $X$ & $M$\\ 
\hline
  $d_{z^2}$@1a & $\Gamma_1(1)$ & $X_1(1)$ & $M_1(1)$\\
  $d_{x^2-y^2}$@1a & $\Gamma_3(1)$ & $X_1(1)$ & $M_3(1)$\\
  $d_{xz,yz}$@2e & $\Gamma_5(2)$ & $X_3(1)\oplus X_4(1) $ & $M_2(1)\oplus M_4(1) $\\
\hline
\hline
\end{tabular*}
\end{center}
\end{table}

{\color{blue}\emph{Material generalization.}}
The key for the $\mathcal{C}=-1$ phase here is rooted in the spin polarized quadratic band touching of $d_{xz},d_{yz}$ orbitals at $\Gamma$ point, where $S_{4z}$ ensures their degeneracy. With SOC and certain orbital occupations, QAH phase can be realized. In fact, the model and analysis from $d$ orbitals above are quite general, and also apply to monolayer Ti$_2$W\textit{X}$_4$ ($X=$ S, Se) with the same lattice structure of $P$-$42m$ symmetry and similar $d$-orbital projected band structure and irreducible representations~\cite{supple}. Ti$_2$W\textit{X}$_4$ has the FM ground state, and the magnetism is from Ti atom. Interestingly, the bandwidth of $d_{z^2}$ orbital is much narrower than that of $d_{xz},d_{yz}$ orbitals in Ti. Each Ti atom is in the $e_g^1t_{2g}^1$ configuration with the magnetic moment close to $2\mu_B$ according to the Hund’s rule. The $e_g$-$t_{2g}$ kinetic exchange leads to strong ferromagnetism~\cite{jiang2023}. Meanwhile, similar two band inversions occur between $d^\uparrow_{xy},d^\uparrow_{yz}$ bands of Ti and $d^\downarrow_{z^2}$ band of W first at $E_F$, and $d^\downarrow_{x^2-y^2}$ band of W then below $E_F$. By adding SOC, the nontrivial topology from quadratic band touching of $d_{xz},d_{yz}$ at $\Gamma$ is now transmitted to the bands below $E_F$, leading to the $\mathcal{C}=-1$ QAH phase (Fig.~\ref{fig2}(g)-(i)). We point out that Ti $d$-orbitals contribute both of topology and magnetism, nevertheless the physics is quite different from KTiSb class of compounds, where the topology is from band inversion from $d_{xz},d_{yz}$ and $d_{z^2}$ at $M$ point~\cite{jiang2023}.

{\color{blue}\emph{Discussions.}}
The topological band structures in these materials have interesting thickness dependence. Bilayer V$_2$WS$_4$ has AA or AB stacking. The magnetic ground state is $A$-type AFM with the out-of-plane easy axis in AB stacking, which is FM within each layer but AFM between the adjacent layer. The $t_{2g}$-$t_{2g}$ exchange of V atoms between neighbor layers via $p$ orbitals of ligand is AFM due to Goodenough-Kanamori-Anderson rule~\cite{khomskii2004}. The system has a full band gap and gapped helical edge state, namely $\mathcal{C}=0$, which can be simply viewed as stacking of two QAH with opposite Chern number along $z$-axis. The FM-$z$ state of bilayer also has a full gap but with $\mathcal{C}=-2$. We further calculate the trilayer case and obtain AFM ground state with an uncompensated FM layer along $z$ axis, and the system is a $\mathcal{C}=-1$ QAH insulator~\cite{supple}. Therefore, we expect $\mathcal{C}$ will oscillate between $-1$ and $0$ depending on odd and even layers of multilayer, as long as it has a full band gap. 

The quantized $\sigma_{xy}$ and vanishing longitudinal conductivity of QAH insulator imply a quantized Kerr/Faraday rotation~\cite{ikebe2010,shimano2013}, when the frequency satisfies $\omega\ll E_g/\hbar$. However, such a quantized magnetooptical effect has not been achieved in existing QAH systems yet due to the small band gap~\cite{okada2016,mogi2022}. Here the topologically nontrivial Chern bands in V$_2$\emph{MX}$_4$ lies far below and above $E_F$, which provides an intriguing but rare platform for exploring giant Kerr rotation from the optical transitions between Chern bands, even possibly in the optical frequency range.

In summary, our work uncover large gap $\mathcal{C}=-1$ QAH phase with interesting interplay between magnetism and topology solely from $d$ orbitals, which applies to a large class of ternary chalcogenide in space group $P$-$42m$. The rich choice of candidate materials indicate the physics is quite general. We hope the theoretical work here could aid the search for new QAH insulators in transition metal compounds.

\begin{acknowledgments}
{\color{blue}\emph{Acknowledgment.}} This work is supported by the National Key Research Program of China under Grant No.~2019YFA0308404, the Natural Science Foundation of China through Grants No.~12350404 and No.~12174066, the Innovation Program for Quantum Science and Technology through Grant No.~2021ZD0302600, the Science and Technology Commission of Shanghai Municipality under Grants No.~23JC1400600 and No.~20JC1415900, Shanghai Municipal Science and Technology Major Project under Grant No.~2019SHZDZX01. Y.J. and H.W. contributed equally to this work.
\end{acknowledgments}


\begin{thebibliography}{79}%
\makeatletter
\providecommand \@ifxundefined [1]{%
 \@ifx{#1\undefined}
}%
\providecommand \@ifnum [1]{%
 \ifnum #1\expandafter \@firstoftwo
 \else \expandafter \@secondoftwo
 \fi
}%
\providecommand \@ifx [1]{%
 \ifx #1\expandafter \@firstoftwo
 \else \expandafter \@secondoftwo
 \fi
}%
\providecommand \natexlab [1]{#1}%
\providecommand \enquote  [1]{``#1''}%
\providecommand \bibnamefont  [1]{#1}%
\providecommand \bibfnamefont [1]{#1}%
\providecommand \citenamefont [1]{#1}%
\providecommand \href@noop [0]{\@secondoftwo}%
\providecommand \href [0]{\begingroup \@sanitize@url \@href}%
\providecommand \@href[1]{\@@startlink{#1}\@@href}%
\providecommand \@@href[1]{\endgroup#1\@@endlink}%
\providecommand \@sanitize@url [0]{\catcode `\\12\catcode `\$12\catcode
  `\&12\catcode `\#12\catcode `\^12\catcode `\_12\catcode `\%12\relax}%
\providecommand \@@startlink[1]{}%
\providecommand \@@endlink[0]{}%
\providecommand \url  [0]{\begingroup\@sanitize@url \@url }%
\providecommand \@url [1]{\endgroup\@href {#1}{\urlprefix }}%
\providecommand \urlprefix  [0]{URL }%
\providecommand \Eprint [0]{\href }%
\providecommand \doibase [0]{http://dx.doi.org/}%
\providecommand \selectlanguage [0]{\@gobble}%
\providecommand \bibinfo  [0]{\@secondoftwo}%
\providecommand \bibfield  [0]{\@secondoftwo}%
\providecommand \translation [1]{[#1]}%
\providecommand \BibitemOpen [0]{}%
\providecommand \bibitemStop [0]{}%
\providecommand \bibitemNoStop [0]{.\EOS\space}%
\providecommand \EOS [0]{\spacefactor3000\relax}%
\providecommand \BibitemShut  [1]{\csname bibitem#1\endcsname}%
\let\auto@bib@innerbib\@empty
%</preamble>
\bibitem [{\citenamefont {Hasan}\ and\ \citenamefont {Kane}(2010)}]{hasan2010}%
  \BibitemOpen
  \bibfield  {author} {\bibinfo {author} {\bibfnamefont {M.~Z.}\ \bibnamefont
  {Hasan}}\ and\ \bibinfo {author} {\bibfnamefont {C.~L.}\ \bibnamefont
  {Kane}},\ }\bibfield  {title} {\enquote {\bibinfo {title}
  {\textit{Colloquium}: Topological insulators},}\ }\href {\doibase
  10.1103/RevModPhys.82.3045} {\bibfield  {journal} {\bibinfo  {journal} {Rev.
  Mod. Phys.}\ }\textbf {\bibinfo {volume} {82}},\ \bibinfo {pages}
  {3045--3067} (\bibinfo {year} {2010})}\BibitemShut {NoStop}%
\bibitem [{\citenamefont {Qi}\ and\ \citenamefont {Zhang}(2011)}]{qi2011}%
  \BibitemOpen
  \bibfield  {author} {\bibinfo {author} {\bibfnamefont {Xiao-Liang}\
  \bibnamefont {Qi}}\ and\ \bibinfo {author} {\bibfnamefont {Shou-Cheng}\
  \bibnamefont {Zhang}},\ }\bibfield  {title} {\enquote {\bibinfo {title}
  {Topological insulators and superconductors},}\ }\href {\doibase
  10.1103/RevModPhys.83.1057} {\bibfield  {journal} {\bibinfo  {journal} {Rev.
  Mod. Phys.}\ }\textbf {\bibinfo {volume} {83}},\ \bibinfo {pages}
  {1057--1110} (\bibinfo {year} {2011})}\BibitemShut {NoStop}%
\bibitem [{\citenamefont {Tokura}\ \emph {et~al.}(2019)\citenamefont {Tokura},
  \citenamefont {Yasuda},\ and\ \citenamefont {Tsukazaki}}]{tokura2019}%
  \BibitemOpen
  \bibfield  {author} {\bibinfo {author} {\bibfnamefont {Yoshinori}\
  \bibnamefont {Tokura}}, \bibinfo {author} {\bibfnamefont {Kenji}\
  \bibnamefont {Yasuda}}, \ and\ \bibinfo {author} {\bibfnamefont {Atsushi}\
  \bibnamefont {Tsukazaki}},\ }\bibfield  {title} {\enquote {\bibinfo {title}
  {Magnetic topological insulators},}\ }\href {\doibase
  10.1038/s42254-018-0011-5} {\bibfield  {journal} {\bibinfo  {journal} {Nature
  Rev. Phys.}\ }\textbf {\bibinfo {volume} {1}},\ \bibinfo {pages} {126--143}
  (\bibinfo {year} {2019})}\BibitemShut {NoStop}%
\bibitem [{\citenamefont {Wang}\ and\ \citenamefont {Zhang}(2017)}]{wang2017c}%
  \BibitemOpen
  \bibfield  {author} {\bibinfo {author} {\bibfnamefont {Jing}\ \bibnamefont
  {Wang}}\ and\ \bibinfo {author} {\bibfnamefont {Shou-Cheng}\ \bibnamefont
  {Zhang}},\ }\bibfield  {title} {\enquote {\bibinfo {title} {Topological
  states of condensed matter},}\ }\href {\doibase 10.1038/NMAT5012} {\bibfield
  {journal} {\bibinfo  {journal} {Nature Mat.}\ }\textbf {\bibinfo {volume}
  {16}},\ \bibinfo {pages} {1062--1067} (\bibinfo {year} {2017})}\BibitemShut
  {NoStop}%
\bibitem [{\citenamefont {Bernevig}\ \emph {et~al.}(2022)\citenamefont
  {Bernevig}, \citenamefont {Felser},\ and\ \citenamefont
  {Beidenkopf}}]{bernevig2022}%
  \BibitemOpen
  \bibfield  {author} {\bibinfo {author} {\bibfnamefont {B.~Andrei}\
  \bibnamefont {Bernevig}}, \bibinfo {author} {\bibfnamefont {Claudia}\
  \bibnamefont {Felser}}, \ and\ \bibinfo {author} {\bibfnamefont {Haim}\
  \bibnamefont {Beidenkopf}},\ }\bibfield  {title} {\enquote {\bibinfo {title}
  {Progress and prospects in magnetic topological materials},}\ }\href
  {\doibase 10.1038/s41586-021-04105-x} {\bibfield  {journal} {\bibinfo
  {journal} {Nature}\ }\textbf {\bibinfo {volume} {603}},\ \bibinfo {pages}
  {41--51} (\bibinfo {year} {2022})}\BibitemShut {NoStop}%
\bibitem [{\citenamefont {Chang}\ \emph {et~al.}(2023)\citenamefont {Chang},
  \citenamefont {Liu},\ and\ \citenamefont {MacDonald}}]{chang2023}%
  \BibitemOpen
  \bibfield  {author} {\bibinfo {author} {\bibfnamefont {Cui-Zu}\ \bibnamefont
  {Chang}}, \bibinfo {author} {\bibfnamefont {Chao-Xing}\ \bibnamefont {Liu}},
  \ and\ \bibinfo {author} {\bibfnamefont {Allan~H.}\ \bibnamefont
  {MacDonald}},\ }\bibfield  {title} {\enquote {\bibinfo {title} {Colloquium:
  Quantum anomalous hall effect},}\ }\href {\doibase
  10.1103/RevModPhys.95.011002} {\bibfield  {journal} {\bibinfo  {journal}
  {Rev. Mod. Phys.}\ }\textbf {\bibinfo {volume} {95}},\ \bibinfo {pages}
  {011002} (\bibinfo {year} {2023})}\BibitemShut {NoStop}%
\bibitem [{\citenamefont {Thouless}\ \emph {et~al.}(1982)\citenamefont
  {Thouless}, \citenamefont {Kohmoto}, \citenamefont {Nightingale},\ and\
  \citenamefont {den Nijs}}]{thouless1982}%
  \BibitemOpen
  \bibfield  {author} {\bibinfo {author} {\bibfnamefont {D.~J.}\ \bibnamefont
  {Thouless}}, \bibinfo {author} {\bibfnamefont {M.}~\bibnamefont {Kohmoto}},
  \bibinfo {author} {\bibfnamefont {M.~P.}\ \bibnamefont {Nightingale}}, \ and\
  \bibinfo {author} {\bibfnamefont {M.}~\bibnamefont {den Nijs}},\ }\bibfield
  {title} {\enquote {\bibinfo {title} {Quantized hall conductance in a
  two-dimensional periodic potential},}\ }\href {\doibase
  10.1103/PhysRevLett.49.405} {\bibfield  {journal} {\bibinfo  {journal} {Phys.
  Rev. Lett.}\ }\textbf {\bibinfo {volume} {49}},\ \bibinfo {pages} {405--408}
  (\bibinfo {year} {1982})}\BibitemShut {NoStop}%
\bibitem [{\citenamefont {Haldane}(1988)}]{haldane1988}%
  \BibitemOpen
  \bibfield  {author} {\bibinfo {author} {\bibfnamefont {F.~D.~M.}\
  \bibnamefont {Haldane}},\ }\bibfield  {title} {\enquote {\bibinfo {title}
  {Model for a quantum hall effect without landau levels: Condensed-matter
  realization of the "parity anomaly"},}\ }\href {\doibase
  10.1103/PhysRevLett.61.2015} {\bibfield  {journal} {\bibinfo  {journal}
  {Phys. Rev. Lett.}\ }\textbf {\bibinfo {volume} {61}},\ \bibinfo {pages}
  {2015--2018} (\bibinfo {year} {1988})}\BibitemShut {NoStop}%
\bibitem [{\citenamefont {Zhang}\ and\ \citenamefont
  {Zhang}(2012)}]{zhang2012}%
  \BibitemOpen
  \bibfield  {author} {\bibinfo {author} {\bibfnamefont {Xiao}\ \bibnamefont
  {Zhang}}\ and\ \bibinfo {author} {\bibfnamefont {Shou-Cheng}\ \bibnamefont
  {Zhang}},\ }\bibfield  {title} {\enquote {\bibinfo {title} {Chiral
  interconnects based on topological insulators},}\ }\href {\doibase
  doi.org/10.1117/12.920325} {\bibfield  {journal} {\bibinfo  {journal} {Proc.
  SPIE Micro- and Nanotechnology Sensors, Systems, and Applications IV}\
  }\textbf {\bibinfo {volume} {8373}},\ \bibinfo {pages} {837309} (\bibinfo
  {year} {2012})}\BibitemShut {NoStop}%
\bibitem [{\citenamefont {Wang}\ \emph {et~al.}(2013)\citenamefont {Wang},
  \citenamefont {Lian}, \citenamefont {Zhang}, \citenamefont {Xu},\ and\
  \citenamefont {Zhang}}]{wang2013a}%
  \BibitemOpen
  \bibfield  {author} {\bibinfo {author} {\bibfnamefont {Jing}\ \bibnamefont
  {Wang}}, \bibinfo {author} {\bibfnamefont {Biao}\ \bibnamefont {Lian}},
  \bibinfo {author} {\bibfnamefont {Haijun}\ \bibnamefont {Zhang}}, \bibinfo
  {author} {\bibfnamefont {Yong}\ \bibnamefont {Xu}}, \ and\ \bibinfo {author}
  {\bibfnamefont {Shou-Cheng}\ \bibnamefont {Zhang}},\ }\bibfield  {title}
  {\enquote {\bibinfo {title} {Quantum anomalous hall effect with higher
  plateaus},}\ }\href {\doibase 10.1103/PhysRevLett.111.136801} {\bibfield
  {journal} {\bibinfo  {journal} {Phys. Rev. Lett.}\ }\textbf {\bibinfo
  {volume} {111}},\ \bibinfo {pages} {136801} (\bibinfo {year}
  {2013})}\BibitemShut {NoStop}%
\bibitem [{\citenamefont {Qi}\ \emph {et~al.}(2010)\citenamefont {Qi},
  \citenamefont {Hughes},\ and\ \citenamefont {Zhang}}]{qi2010b}%
  \BibitemOpen
  \bibfield  {author} {\bibinfo {author} {\bibfnamefont {Xiao-Liang}\
  \bibnamefont {Qi}}, \bibinfo {author} {\bibfnamefont {Taylor~L.}\
  \bibnamefont {Hughes}}, \ and\ \bibinfo {author} {\bibfnamefont {Shou-Cheng}\
  \bibnamefont {Zhang}},\ }\bibfield  {title} {\enquote {\bibinfo {title}
  {Chiral topological superconductor from the quantum hall state},}\ }\href
  {\doibase 10.1103/PhysRevB.82.184516} {\bibfield  {journal} {\bibinfo
  {journal} {Phys. Rev. B}\ }\textbf {\bibinfo {volume} {82}},\ \bibinfo
  {pages} {184516} (\bibinfo {year} {2010})}\BibitemShut {NoStop}%
\bibitem [{\citenamefont {Wang}\ \emph
  {et~al.}(2015{\natexlab{a}})\citenamefont {Wang}, \citenamefont {Zhou},
  \citenamefont {Lian},\ and\ \citenamefont {Zhang}}]{wang2015c}%
  \BibitemOpen
  \bibfield  {author} {\bibinfo {author} {\bibfnamefont {Jing}\ \bibnamefont
  {Wang}}, \bibinfo {author} {\bibfnamefont {Quan}\ \bibnamefont {Zhou}},
  \bibinfo {author} {\bibfnamefont {Biao}\ \bibnamefont {Lian}}, \ and\
  \bibinfo {author} {\bibfnamefont {Shou-Cheng}\ \bibnamefont {Zhang}},\
  }\bibfield  {title} {\enquote {\bibinfo {title} {Chiral topological
  superconductor and half-integer conductance plateau from quantum anomalous
  hall plateau transition},}\ }\href {\doibase 10.1103/PhysRevB.92.064520}
  {\bibfield  {journal} {\bibinfo  {journal} {Phys. Rev. B}\ }\textbf {\bibinfo
  {volume} {92}},\ \bibinfo {pages} {064520} (\bibinfo {year}
  {2015}{\natexlab{a}})}\BibitemShut {NoStop}%
\bibitem [{\citenamefont {Lian}\ \emph {et~al.}(2018)\citenamefont {Lian},
  \citenamefont {Sun}, \citenamefont {Vaezi}, \citenamefont {Qi},\ and\
  \citenamefont {Zhang}}]{lian2018b}%
  \BibitemOpen
  \bibfield  {author} {\bibinfo {author} {\bibfnamefont {Biao}\ \bibnamefont
  {Lian}}, \bibinfo {author} {\bibfnamefont {Xiao-Qi}\ \bibnamefont {Sun}},
  \bibinfo {author} {\bibfnamefont {Abolhassan}\ \bibnamefont {Vaezi}},
  \bibinfo {author} {\bibfnamefont {Xiao-Liang}\ \bibnamefont {Qi}}, \ and\
  \bibinfo {author} {\bibfnamefont {Shou-Cheng}\ \bibnamefont {Zhang}},\
  }\bibfield  {title} {\enquote {\bibinfo {title} {Topological quantum
  computation based on chiral majorana fermions},}\ }\href {\doibase
  10.1073/pnas.1810003115} {\bibfield  {journal} {\bibinfo  {journal} {Proc.
  Natl. Acad. Sci. USA}\ }\textbf {\bibinfo {volume} {115}},\ \bibinfo {pages}
  {10938--10942} (\bibinfo {year} {2018})}\BibitemShut {NoStop}%
\bibitem [{\citenamefont {Chang}\ \emph {et~al.}(2013)\citenamefont {Chang},
  \citenamefont {Zhang}, \citenamefont {Feng}, \citenamefont {Shen},
  \citenamefont {Zhang}, \citenamefont {Guo}, \citenamefont {Li}, \citenamefont
  {Ou}, \citenamefont {Wei}, \citenamefont {Wang}, \citenamefont {Ji},
  \citenamefont {Feng}, \citenamefont {Ji}, \citenamefont {Chen}, \citenamefont
  {Jia}, \citenamefont {Dai}, \citenamefont {Fang}, \citenamefont {Zhang},
  \citenamefont {He}, \citenamefont {Wang}, \citenamefont {Lu}, \citenamefont
  {Ma},\ and\ \citenamefont {Xue}}]{chang2013b}%
  \BibitemOpen
  \bibfield  {author} {\bibinfo {author} {\bibfnamefont {Cui-Zu}\ \bibnamefont
  {Chang}}, \bibinfo {author} {\bibfnamefont {Jinsong}\ \bibnamefont {Zhang}},
  \bibinfo {author} {\bibfnamefont {Xiao}\ \bibnamefont {Feng}}, \bibinfo
  {author} {\bibfnamefont {Jie}\ \bibnamefont {Shen}}, \bibinfo {author}
  {\bibfnamefont {Zuocheng}\ \bibnamefont {Zhang}}, \bibinfo {author}
  {\bibfnamefont {Minghua}\ \bibnamefont {Guo}}, \bibinfo {author}
  {\bibfnamefont {Kang}\ \bibnamefont {Li}}, \bibinfo {author} {\bibfnamefont
  {Yunbo}\ \bibnamefont {Ou}}, \bibinfo {author} {\bibfnamefont {Pang}\
  \bibnamefont {Wei}}, \bibinfo {author} {\bibfnamefont {Li-Li}\ \bibnamefont
  {Wang}}, \bibinfo {author} {\bibfnamefont {Zhong-Qing}\ \bibnamefont {Ji}},
  \bibinfo {author} {\bibfnamefont {Yang}\ \bibnamefont {Feng}}, \bibinfo
  {author} {\bibfnamefont {Shuaihua}\ \bibnamefont {Ji}}, \bibinfo {author}
  {\bibfnamefont {Xi}~\bibnamefont {Chen}}, \bibinfo {author} {\bibfnamefont
  {Jinfeng}\ \bibnamefont {Jia}}, \bibinfo {author} {\bibfnamefont
  {Xi}~\bibnamefont {Dai}}, \bibinfo {author} {\bibfnamefont {Zhong}\
  \bibnamefont {Fang}}, \bibinfo {author} {\bibfnamefont {Shou-Cheng}\
  \bibnamefont {Zhang}}, \bibinfo {author} {\bibfnamefont {Ke}~\bibnamefont
  {He}}, \bibinfo {author} {\bibfnamefont {Yayu}\ \bibnamefont {Wang}},
  \bibinfo {author} {\bibfnamefont {Li}~\bibnamefont {Lu}}, \bibinfo {author}
  {\bibfnamefont {Xu-Cun}\ \bibnamefont {Ma}}, \ and\ \bibinfo {author}
  {\bibfnamefont {Qi-Kun}\ \bibnamefont {Xue}},\ }\bibfield  {title} {\enquote
  {\bibinfo {title} {{Experimental Observation of the Quantum Anomalous Hall
  Effect in a Magnetic Topological Insulator}},}\ }\href {\doibase
  10.1126/science.1234414} {\bibfield  {journal} {\bibinfo  {journal}
  {Science}\ }\textbf {\bibinfo {volume} {340}},\ \bibinfo {pages} {167--170}
  (\bibinfo {year} {2013})}\BibitemShut {NoStop}%
\bibitem [{\citenamefont {Chang}\ \emph {et~al.}(2015)\citenamefont {Chang},
  \citenamefont {Zhao}, \citenamefont {Kim}, \citenamefont {Zhang},
  \citenamefont {Assaf}, \citenamefont {Heiman}, \citenamefont {Zhang},
  \citenamefont {Liu}, \citenamefont {Chan},\ and\ \citenamefont
  {Moodera}}]{chang2015}%
  \BibitemOpen
  \bibfield  {author} {\bibinfo {author} {\bibfnamefont {Cui-Zu}\ \bibnamefont
  {Chang}}, \bibinfo {author} {\bibfnamefont {Weiwei}\ \bibnamefont {Zhao}},
  \bibinfo {author} {\bibfnamefont {Duk~Y.}\ \bibnamefont {Kim}}, \bibinfo
  {author} {\bibfnamefont {Haijun}\ \bibnamefont {Zhang}}, \bibinfo {author}
  {\bibfnamefont {Badih~A.}\ \bibnamefont {Assaf}}, \bibinfo {author}
  {\bibfnamefont {Don}\ \bibnamefont {Heiman}}, \bibinfo {author}
  {\bibfnamefont {Shou-Cheng}\ \bibnamefont {Zhang}}, \bibinfo {author}
  {\bibfnamefont {Chaoxing}\ \bibnamefont {Liu}}, \bibinfo {author}
  {\bibfnamefont {Moses H.~W.}\ \bibnamefont {Chan}}, \ and\ \bibinfo {author}
  {\bibfnamefont {Jagadeesh~S.}\ \bibnamefont {Moodera}},\ }\bibfield  {title}
  {\enquote {\bibinfo {title} {High-precision realization of robust quantum
  anomalous hall state in a hard ferromagnetic topological insulator},}\ }\href
  {\doibase 10.1038/nmat4204} {\bibfield  {journal} {\bibinfo  {journal}
  {Nature Mater.}\ }\textbf {\bibinfo {volume} {14}},\ \bibinfo {pages} {473}
  (\bibinfo {year} {2015})}\BibitemShut {NoStop}%
\bibitem [{\citenamefont {Mogi}\ \emph {et~al.}(2015)\citenamefont {Mogi},
  \citenamefont {Yoshimi}, \citenamefont {Tsukazaki}, \citenamefont {Yasuda},
  \citenamefont {Kozuka}, \citenamefont {Takahashi}, \citenamefont {Kawasaki},\
  and\ \citenamefont {Tokura}}]{mogi2015}%
  \BibitemOpen
  \bibfield  {author} {\bibinfo {author} {\bibfnamefont {M.}~\bibnamefont
  {Mogi}}, \bibinfo {author} {\bibfnamefont {R.}~\bibnamefont {Yoshimi}},
  \bibinfo {author} {\bibfnamefont {A.}~\bibnamefont {Tsukazaki}}, \bibinfo
  {author} {\bibfnamefont {K.}~\bibnamefont {Yasuda}}, \bibinfo {author}
  {\bibfnamefont {Y.}~\bibnamefont {Kozuka}}, \bibinfo {author} {\bibfnamefont
  {K.~S.}\ \bibnamefont {Takahashi}}, \bibinfo {author} {\bibfnamefont
  {M.}~\bibnamefont {Kawasaki}}, \ and\ \bibinfo {author} {\bibfnamefont
  {Y.}~\bibnamefont {Tokura}},\ }\bibfield  {title} {\enquote {\bibinfo {title}
  {Magnetic modulation doping in topological insulators toward
  higher-temperature quantum anomalous hall effect},}\ }\href {\doibase
  10.1063/1.4935075} {\bibfield  {journal} {\bibinfo  {journal} {Appl. Phys.
  Lett.}\ }\textbf {\bibinfo {volume} {107}},\ \bibinfo {pages} {182401}
  (\bibinfo {year} {2015})}\BibitemShut {NoStop}%
\bibitem [{\citenamefont {Bestwick}\ \emph {et~al.}(2015)\citenamefont
  {Bestwick}, \citenamefont {Fox}, \citenamefont {Kou}, \citenamefont {Pan},
  \citenamefont {Wang},\ and\ \citenamefont {Goldhaber-Gordon}}]{bestwick2015}%
  \BibitemOpen
  \bibfield  {author} {\bibinfo {author} {\bibfnamefont {A.~J.}\ \bibnamefont
  {Bestwick}}, \bibinfo {author} {\bibfnamefont {E.~J.}\ \bibnamefont {Fox}},
  \bibinfo {author} {\bibfnamefont {Xufeng}\ \bibnamefont {Kou}}, \bibinfo
  {author} {\bibfnamefont {Lei}\ \bibnamefont {Pan}}, \bibinfo {author}
  {\bibfnamefont {Kang~L.}\ \bibnamefont {Wang}}, \ and\ \bibinfo {author}
  {\bibfnamefont {D.}~\bibnamefont {Goldhaber-Gordon}},\ }\bibfield  {title}
  {\enquote {\bibinfo {title} {Precise quantization of the anomalous hall
  effect near zero magnetic field},}\ }\href {\doibase
  10.1103/PhysRevLett.114.187201} {\bibfield  {journal} {\bibinfo  {journal}
  {Phys. Rev. Lett.}\ }\textbf {\bibinfo {volume} {114}},\ \bibinfo {pages}
  {187201} (\bibinfo {year} {2015})}\BibitemShut {NoStop}%
\bibitem [{\citenamefont {Watanabe}\ \emph {et~al.}(2019)\citenamefont
  {Watanabe}, \citenamefont {Yoshimi}, \citenamefont {Kawamura}, \citenamefont
  {Mogi}, \citenamefont {Tsukazaki}, \citenamefont {Yu}, \citenamefont
  {Nakajima}, \citenamefont {Takahashi}, \citenamefont {Kawasaki},\ and\
  \citenamefont {Tokura}}]{watanabe2019}%
  \BibitemOpen
  \bibfield  {author} {\bibinfo {author} {\bibfnamefont {R.}~\bibnamefont
  {Watanabe}}, \bibinfo {author} {\bibfnamefont {R.}~\bibnamefont {Yoshimi}},
  \bibinfo {author} {\bibfnamefont {M.}~\bibnamefont {Kawamura}}, \bibinfo
  {author} {\bibfnamefont {M.}~\bibnamefont {Mogi}}, \bibinfo {author}
  {\bibfnamefont {A.}~\bibnamefont {Tsukazaki}}, \bibinfo {author}
  {\bibfnamefont {X.~Z.}\ \bibnamefont {Yu}}, \bibinfo {author} {\bibfnamefont
  {K.}~\bibnamefont {Nakajima}}, \bibinfo {author} {\bibfnamefont {K.~S.}\
  \bibnamefont {Takahashi}}, \bibinfo {author} {\bibfnamefont {M.}~\bibnamefont
  {Kawasaki}}, \ and\ \bibinfo {author} {\bibfnamefont {Y.}~\bibnamefont
  {Tokura}},\ }\bibfield  {title} {\enquote {\bibinfo {title} {Quantum
  anomalous hall effect driven by magnetic proximity coupling in all-telluride
  based heterostructure},}\ }\href {\doibase 10.1063/1.5111891} {\bibfield
  {journal} {\bibinfo  {journal} {Appl. Phys. Lett.}\ }\textbf {\bibinfo
  {volume} {115}},\ \bibinfo {pages} {102403} (\bibinfo {year}
  {2019})}\BibitemShut {NoStop}%
\bibitem [{\citenamefont {Deng}\ \emph {et~al.}(2020)\citenamefont {Deng},
  \citenamefont {Yu}, \citenamefont {Shi}, \citenamefont {Guo}, \citenamefont
  {Xu}, \citenamefont {Wang}, \citenamefont {Chen},\ and\ \citenamefont
  {Zhang}}]{deng2020}%
  \BibitemOpen
  \bibfield  {author} {\bibinfo {author} {\bibfnamefont {Yujun}\ \bibnamefont
  {Deng}}, \bibinfo {author} {\bibfnamefont {Yijun}\ \bibnamefont {Yu}},
  \bibinfo {author} {\bibfnamefont {Meng~Zhu}\ \bibnamefont {Shi}}, \bibinfo
  {author} {\bibfnamefont {Zhongxun}\ \bibnamefont {Guo}}, \bibinfo {author}
  {\bibfnamefont {Zihan}\ \bibnamefont {Xu}}, \bibinfo {author} {\bibfnamefont
  {Jing}\ \bibnamefont {Wang}}, \bibinfo {author} {\bibfnamefont {Xian~Hui}\
  \bibnamefont {Chen}}, \ and\ \bibinfo {author} {\bibfnamefont {Yuanbo}\
  \bibnamefont {Zhang}},\ }\bibfield  {title} {\enquote {\bibinfo {title}
  {Quantum anomalous hall effect in intrinsic magnetic topological insulator
  mnbi2te4},}\ }\href {\doibase 10.1126/science.aax8156} {\bibfield  {journal}
  {\bibinfo  {journal} {Science}\ }\textbf {\bibinfo {volume} {367}},\ \bibinfo
  {pages} {895--900} (\bibinfo {year} {2020})}\BibitemShut {NoStop}%
\bibitem [{\citenamefont {Serlin}\ \emph {et~al.}(2020)\citenamefont {Serlin},
  \citenamefont {Tschirhart}, \citenamefont {Polshyn}, \citenamefont {Zhang},
  \citenamefont {Zhu}, \citenamefont {Watanabe}, \citenamefont {Taniguchi},
  \citenamefont {Balents},\ and\ \citenamefont {Young}}]{serlin2020}%
  \BibitemOpen
  \bibfield  {author} {\bibinfo {author} {\bibfnamefont {M.}~\bibnamefont
  {Serlin}}, \bibinfo {author} {\bibfnamefont {C.~L.}\ \bibnamefont
  {Tschirhart}}, \bibinfo {author} {\bibfnamefont {H.}~\bibnamefont {Polshyn}},
  \bibinfo {author} {\bibfnamefont {Y.}~\bibnamefont {Zhang}}, \bibinfo
  {author} {\bibfnamefont {J.}~\bibnamefont {Zhu}}, \bibinfo {author}
  {\bibfnamefont {K.}~\bibnamefont {Watanabe}}, \bibinfo {author}
  {\bibfnamefont {T.}~\bibnamefont {Taniguchi}}, \bibinfo {author}
  {\bibfnamefont {L.}~\bibnamefont {Balents}}, \ and\ \bibinfo {author}
  {\bibfnamefont {A.~F.}\ \bibnamefont {Young}},\ }\bibfield  {title} {\enquote
  {\bibinfo {title} {Intrinsic quantized anomalous hall effect in a moir\'e
  heterostructure},}\ }\href {\doibase 10.1126/science.aay5533} {\bibfield
  {journal} {\bibinfo  {journal} {Science}\ }\textbf {\bibinfo {volume}
  {367}},\ \bibinfo {pages} {900--903} (\bibinfo {year} {2020})}\BibitemShut
  {NoStop}%
\bibitem [{\citenamefont {Li}\ \emph {et~al.}(2021)\citenamefont {Li},
  \citenamefont {Jiang}, \citenamefont {Shen}, \citenamefont {Zhang},
  \citenamefont {Li}, \citenamefont {Tao}, \citenamefont {Devakul},
  \citenamefont {Watanabe}, \citenamefont {Taniguchi}, \citenamefont {Fu},
  \citenamefont {Shan},\ and\ \citenamefont {Mak}}]{li2021}%
  \BibitemOpen
  \bibfield  {author} {\bibinfo {author} {\bibfnamefont {Tingxin}\ \bibnamefont
  {Li}}, \bibinfo {author} {\bibfnamefont {Shengwei}\ \bibnamefont {Jiang}},
  \bibinfo {author} {\bibfnamefont {Bowen}\ \bibnamefont {Shen}}, \bibinfo
  {author} {\bibfnamefont {Yang}\ \bibnamefont {Zhang}}, \bibinfo {author}
  {\bibfnamefont {Lizhong}\ \bibnamefont {Li}}, \bibinfo {author}
  {\bibfnamefont {Zui}\ \bibnamefont {Tao}}, \bibinfo {author} {\bibfnamefont
  {Trithep}\ \bibnamefont {Devakul}}, \bibinfo {author} {\bibfnamefont {Kenji}\
  \bibnamefont {Watanabe}}, \bibinfo {author} {\bibfnamefont {Takashi}\
  \bibnamefont {Taniguchi}}, \bibinfo {author} {\bibfnamefont {Liang}\
  \bibnamefont {Fu}}, \bibinfo {author} {\bibfnamefont {Jie}\ \bibnamefont
  {Shan}}, \ and\ \bibinfo {author} {\bibfnamefont {Kin~Fai}\ \bibnamefont
  {Mak}},\ }\bibfield  {title} {\enquote {\bibinfo {title} {Quantum anomalous
  hall effect from intertwined moir\'e bands},}\ }\href {\doibase
  10.1038/s41586-021-04171-1} {\bibfield  {journal} {\bibinfo  {journal}
  {Nature}\ }\textbf {\bibinfo {volume} {600}},\ \bibinfo {pages} {641--646}
  (\bibinfo {year} {2021})}\BibitemShut {NoStop}%
\bibitem [{\citenamefont {Okazaki}\ \emph {et~al.}(2022)\citenamefont
  {Okazaki}, \citenamefont {Oe}, \citenamefont {Kawamura}, \citenamefont
  {Yoshimi}, \citenamefont {Nakamura}, \citenamefont {Takada}, \citenamefont
  {Mogi}, \citenamefont {Takahashi}, \citenamefont {Tsukazaki}, \citenamefont
  {Kawasaki}, \citenamefont {Tokura},\ and\ \citenamefont
  {Kaneko}}]{okazaki2022}%
  \BibitemOpen
  \bibfield  {author} {\bibinfo {author} {\bibfnamefont {Yuma}\ \bibnamefont
  {Okazaki}}, \bibinfo {author} {\bibfnamefont {Takehiko}\ \bibnamefont {Oe}},
  \bibinfo {author} {\bibfnamefont {Minoru}\ \bibnamefont {Kawamura}}, \bibinfo
  {author} {\bibfnamefont {Ryutaro}\ \bibnamefont {Yoshimi}}, \bibinfo {author}
  {\bibfnamefont {Shuji}\ \bibnamefont {Nakamura}}, \bibinfo {author}
  {\bibfnamefont {Shintaro}\ \bibnamefont {Takada}}, \bibinfo {author}
  {\bibfnamefont {Masataka}\ \bibnamefont {Mogi}}, \bibinfo {author}
  {\bibfnamefont {Kei~S.}\ \bibnamefont {Takahashi}}, \bibinfo {author}
  {\bibfnamefont {Atsushi}\ \bibnamefont {Tsukazaki}}, \bibinfo {author}
  {\bibfnamefont {Masashi}\ \bibnamefont {Kawasaki}}, \bibinfo {author}
  {\bibfnamefont {Yoshinori}\ \bibnamefont {Tokura}}, \ and\ \bibinfo {author}
  {\bibfnamefont {Nobu-Hisa}\ \bibnamefont {Kaneko}},\ }\bibfield  {title}
  {\enquote {\bibinfo {title} {Quantum anomalous hall effect with a permanent
  magnet defines a quantum resistance standard},}\ }\href {\doibase
  10.1038/s41567-021-01424-8} {\bibfield  {journal} {\bibinfo  {journal}
  {Nature Phys.}\ }\textbf {\bibinfo {volume} {18}},\ \bibinfo {pages} {25--29}
  (\bibinfo {year} {2022})}\BibitemShut {NoStop}%
\bibitem [{\citenamefont {You}\ \emph {et~al.}(2019)\citenamefont {You},
  \citenamefont {Zhang}, \citenamefont {Gu},\ and\ \citenamefont
  {Su}}]{you2019}%
  \BibitemOpen
  \bibfield  {author} {\bibinfo {author} {\bibfnamefont {Jing-Yang}\
  \bibnamefont {You}}, \bibinfo {author} {\bibfnamefont {Zhen}\ \bibnamefont
  {Zhang}}, \bibinfo {author} {\bibfnamefont {Bo}~\bibnamefont {Gu}}, \ and\
  \bibinfo {author} {\bibfnamefont {Gang}\ \bibnamefont {Su}},\ }\bibfield
  {title} {\enquote {\bibinfo {title} {Two-dimensional room-temperature
  ferromagnetic semiconductors with quantum anomalous hall effect},}\ }\href
  {\doibase 10.1103/PhysRevApplied.12.024063} {\bibfield  {journal} {\bibinfo
  {journal} {Phys. Rev. Applied}\ }\textbf {\bibinfo {volume} {12}},\ \bibinfo
  {pages} {024063} (\bibinfo {year} {2019})}\BibitemShut {NoStop}%
\bibitem [{\citenamefont {Sun}\ \emph {et~al.}(2020{\natexlab{a}})\citenamefont
  {Sun}, \citenamefont {Zhong}, \citenamefont {Cui}, \citenamefont {Shi},
  \citenamefont {Hao}, \citenamefont {Xu},\ and\ \citenamefont
  {Li}}]{sunj2020}%
  \BibitemOpen
  \bibfield  {author} {\bibinfo {author} {\bibfnamefont {Jiaxiang}\
  \bibnamefont {Sun}}, \bibinfo {author} {\bibfnamefont {Xin}\ \bibnamefont
  {Zhong}}, \bibinfo {author} {\bibfnamefont {Wenwen}\ \bibnamefont {Cui}},
  \bibinfo {author} {\bibfnamefont {Jingming}\ \bibnamefont {Shi}}, \bibinfo
  {author} {\bibfnamefont {Jian}\ \bibnamefont {Hao}}, \bibinfo {author}
  {\bibfnamefont {Meiling}\ \bibnamefont {Xu}}, \ and\ \bibinfo {author}
  {\bibfnamefont {Yinwei}\ \bibnamefont {Li}},\ }\bibfield  {title} {\enquote
  {\bibinfo {title} {The intrinsic magnetism{,} quantum anomalous hall effect
  and curie temperature in 2d transition metal trihalides},}\ }\href {\doibase
  10.1039/C9CP05084A} {\bibfield  {journal} {\bibinfo  {journal} {Phys. Chem.
  Chem. Phys.}\ }\textbf {\bibinfo {volume} {22}},\ \bibinfo {pages}
  {2429--2436} (\bibinfo {year} {2020}{\natexlab{a}})}\BibitemShut {NoStop}%
\bibitem [{\citenamefont {Li}\ \emph {et~al.}(2020)\citenamefont {Li},
  \citenamefont {Li}, \citenamefont {Li}, \citenamefont {Ye}, \citenamefont
  {Zheng}, \citenamefont {Zhang}, \citenamefont {Fu}, \citenamefont {Duan},\
  and\ \citenamefont {Xu}}]{liy2020}%
  \BibitemOpen
  \bibfield  {author} {\bibinfo {author} {\bibfnamefont {Yang}\ \bibnamefont
  {Li}}, \bibinfo {author} {\bibfnamefont {Jiaheng}\ \bibnamefont {Li}},
  \bibinfo {author} {\bibfnamefont {Yang}\ \bibnamefont {Li}}, \bibinfo
  {author} {\bibfnamefont {Meng}\ \bibnamefont {Ye}}, \bibinfo {author}
  {\bibfnamefont {Fawei}\ \bibnamefont {Zheng}}, \bibinfo {author}
  {\bibfnamefont {Zetao}\ \bibnamefont {Zhang}}, \bibinfo {author}
  {\bibfnamefont {Jingheng}\ \bibnamefont {Fu}}, \bibinfo {author}
  {\bibfnamefont {Wenhui}\ \bibnamefont {Duan}}, \ and\ \bibinfo {author}
  {\bibfnamefont {Yong}\ \bibnamefont {Xu}},\ }\bibfield  {title} {\enquote
  {\bibinfo {title} {High-temperature quantum anomalous hall insulators in
  lithium-decorated iron-based superconductor materials},}\ }\href {\doibase
  10.1103/PhysRevLett.125.086401} {\bibfield  {journal} {\bibinfo  {journal}
  {Phys. Rev. Lett.}\ }\textbf {\bibinfo {volume} {125}},\ \bibinfo {pages}
  {086401} (\bibinfo {year} {2020})}\BibitemShut {NoStop}%
\bibitem [{\citenamefont {Xuan}\ \emph {et~al.}(2022)\citenamefont {Xuan},
  \citenamefont {Zhang}, \citenamefont {Chen},\ and\ \citenamefont
  {Guo}}]{xuan2022}%
  \BibitemOpen
  \bibfield  {author} {\bibinfo {author} {\bibfnamefont {X.}~\bibnamefont
  {Xuan}}, \bibinfo {author} {\bibfnamefont {Z.}~\bibnamefont {Zhang}},
  \bibinfo {author} {\bibfnamefont {C.}~\bibnamefont {Chen}}, \ and\ \bibinfo
  {author} {\bibfnamefont {W.}~\bibnamefont {Guo}},\ }\bibfield  {title}
  {\enquote {\bibinfo {title} {Robust quantum anomalous hall states in
  monolayer and few-layer tite},}\ }\href {\doibase
  10.1021/acs.nanolett.2c01421} {\bibfield  {journal} {\bibinfo  {journal}
  {Nano Lett.}\ }\textbf {\bibinfo {volume} {22}},\ \bibinfo {pages}
  {5379--5384} (\bibinfo {year} {2022})}\BibitemShut {NoStop}%
\bibitem [{\citenamefont {Sun}\ \emph {et~al.}(2020{\natexlab{b}})\citenamefont
  {Sun}, \citenamefont {Ma},\ and\ \citenamefont {Kioussis}}]{sun2020}%
  \BibitemOpen
  \bibfield  {author} {\bibinfo {author} {\bibfnamefont {Qilong}\ \bibnamefont
  {Sun}}, \bibinfo {author} {\bibfnamefont {Yandong}\ \bibnamefont {Ma}}, \
  and\ \bibinfo {author} {\bibfnamefont {Nicholas}\ \bibnamefont {Kioussis}},\
  }\bibfield  {title} {\enquote {\bibinfo {title} {Two-dimensional dirac
  spin-gapless semiconductors with tunable perpendicular magnetic anisotropy
  and a robust quantum anomalous hall effect},}\ }\href {\doibase
  10.1039/d0mh00396d} {\bibfield  {journal} {\bibinfo  {journal} {Mater.
  Horiz}\ }\textbf {\bibinfo {volume} {7}},\ \bibinfo {pages} {2071--2077}
  (\bibinfo {year} {2020}{\natexlab{b}})}\BibitemShut {NoStop}%
\bibitem [{\citenamefont {Li}\ \emph {et~al.}(2022)\citenamefont {Li},
  \citenamefont {Han},\ and\ \citenamefont {Qiao}}]{li2022}%
  \BibitemOpen
  \bibfield  {author} {\bibinfo {author} {\bibfnamefont {Zeyu}\ \bibnamefont
  {Li}}, \bibinfo {author} {\bibfnamefont {Yulei}\ \bibnamefont {Han}}, \ and\
  \bibinfo {author} {\bibfnamefont {Zhenhua}\ \bibnamefont {Qiao}},\ }\bibfield
   {title} {\enquote {\bibinfo {title} {Chern number tunable quantum anomalous
  hall effect in monolayer transitional metal oxides via manipulating
  magnetization orientation},}\ }\href {\doibase
  10.1103/PhysRevLett.129.036801} {\bibfield  {journal} {\bibinfo  {journal}
  {Phys. Rev. Lett.}\ }\textbf {\bibinfo {volume} {129}},\ \bibinfo {pages}
  {036801} (\bibinfo {year} {2022})}\BibitemShut {NoStop}%
\bibitem [{\citenamefont {Jiang}\ \emph {et~al.}(2023)\citenamefont {Jiang},
  \citenamefont {Wang},\ and\ \citenamefont {Wang}}]{jiang2023}%
  \BibitemOpen
  \bibfield  {author} {\bibinfo {author} {\bibfnamefont {Yadong}\ \bibnamefont
  {Jiang}}, \bibinfo {author} {\bibfnamefont {Huan}\ \bibnamefont {Wang}}, \
  and\ \bibinfo {author} {\bibfnamefont {Jing}\ \bibnamefont {Wang}},\
  }\bibfield  {title} {\enquote {\bibinfo {title} {Large-gap quantum anomalous
  hall insulators in the $a\mathrm{Ti}x$ ($a=\mathrm{K}$, rb, sr; $x$=sb, bi,
  sn) class of compounds},}\ }\href {\doibase 10.1103/PhysRevB.108.165122}
  {\bibfield  {journal} {\bibinfo  {journal} {Phys. Rev. B}\ }\textbf {\bibinfo
  {volume} {108}},\ \bibinfo {pages} {165122} (\bibinfo {year}
  {2023})}\BibitemShut {NoStop}%
\bibitem [{\citenamefont {Liu}\ \emph {et~al.}(2008)\citenamefont {Liu},
  \citenamefont {Qi}, \citenamefont {Dai}, \citenamefont {Fang},\ and\
  \citenamefont {Zhang}}]{liu2008}%
  \BibitemOpen
  \bibfield  {author} {\bibinfo {author} {\bibfnamefont {Chao-Xing}\
  \bibnamefont {Liu}}, \bibinfo {author} {\bibfnamefont {Xiao-Liang}\
  \bibnamefont {Qi}}, \bibinfo {author} {\bibfnamefont {Xi}~\bibnamefont
  {Dai}}, \bibinfo {author} {\bibfnamefont {Zhong}\ \bibnamefont {Fang}}, \
  and\ \bibinfo {author} {\bibfnamefont {Shou-Cheng}\ \bibnamefont {Zhang}},\
  }\bibfield  {title} {\enquote {\bibinfo {title} {Quantum anomalous hall
  effect in $\mathrm{Hg}_{1-y}\mathrm{Mn}_{y}\mathrm{Te}$ quantum wells},}\
  }\href {\doibase 10.1103/PhysRevLett.101.146802} {\bibfield  {journal}
  {\bibinfo  {journal} {Phys. Rev. Lett.}\ }\textbf {\bibinfo {volume} {101}},\
  \bibinfo {pages} {146802} (\bibinfo {year} {2008})}\BibitemShut {NoStop}%
\bibitem [{\citenamefont {Yu}\ \emph {et~al.}(2010)\citenamefont {Yu},
  \citenamefont {Zhang}, \citenamefont {Zhang}, \citenamefont {Zhang},
  \citenamefont {Dai},\ and\ \citenamefont {Fang}}]{yu2010}%
  \BibitemOpen
  \bibfield  {author} {\bibinfo {author} {\bibfnamefont {Rui}\ \bibnamefont
  {Yu}}, \bibinfo {author} {\bibfnamefont {Wei}\ \bibnamefont {Zhang}},
  \bibinfo {author} {\bibfnamefont {Hai-Jun}\ \bibnamefont {Zhang}}, \bibinfo
  {author} {\bibfnamefont {Shou-Cheng}\ \bibnamefont {Zhang}}, \bibinfo
  {author} {\bibfnamefont {Xi}~\bibnamefont {Dai}}, \ and\ \bibinfo {author}
  {\bibfnamefont {Zhong}\ \bibnamefont {Fang}},\ }\bibfield  {title} {\enquote
  {\bibinfo {title} {{Quantized Anomalous Hall Effect in Magnetic Topological
  Insulators}},}\ }\href {\doibase 10.1126/science.1187485} {\bibfield
  {journal} {\bibinfo  {journal} {Science}\ }\textbf {\bibinfo {volume}
  {329}},\ \bibinfo {pages} {61--64} (\bibinfo {year} {2010})}\BibitemShut
  {NoStop}%
\bibitem [{\citenamefont {Wang}\ \emph
  {et~al.}(2015{\natexlab{b}})\citenamefont {Wang}, \citenamefont {Lian},\ and\
  \citenamefont {Zhang}}]{wang2015d}%
  \BibitemOpen
  \bibfield  {author} {\bibinfo {author} {\bibfnamefont {Jing}\ \bibnamefont
  {Wang}}, \bibinfo {author} {\bibfnamefont {Biao}\ \bibnamefont {Lian}}, \
  and\ \bibinfo {author} {\bibfnamefont {Shou-Cheng}\ \bibnamefont {Zhang}},\
  }\bibfield  {title} {\enquote {\bibinfo {title} {Quantum anomalous hall
  effect in magnetic topological insulators},}\ }\href {\doibase
  10.1088/0031-8949/2015/T164/014003} {\bibfield  {journal} {\bibinfo
  {journal} {Phys. Scr.}\ }\textbf {\bibinfo {volume} {T164}},\ \bibinfo
  {pages} {014003} (\bibinfo {year} {2015}{\natexlab{b}})}\BibitemShut
  {NoStop}%
\bibitem [{\citenamefont {Zhang}\ \emph {et~al.}(2019)\citenamefont {Zhang},
  \citenamefont {Shi}, \citenamefont {Zhu}, \citenamefont {Xing}, \citenamefont
  {Zhang},\ and\ \citenamefont {Wang}}]{zhang2019}%
  \BibitemOpen
  \bibfield  {author} {\bibinfo {author} {\bibfnamefont {Dongqin}\ \bibnamefont
  {Zhang}}, \bibinfo {author} {\bibfnamefont {Minji}\ \bibnamefont {Shi}},
  \bibinfo {author} {\bibfnamefont {Tongshuai}\ \bibnamefont {Zhu}}, \bibinfo
  {author} {\bibfnamefont {Dingyu}\ \bibnamefont {Xing}}, \bibinfo {author}
  {\bibfnamefont {Haijun}\ \bibnamefont {Zhang}}, \ and\ \bibinfo {author}
  {\bibfnamefont {Jing}\ \bibnamefont {Wang}},\ }\bibfield  {title} {\enquote
  {\bibinfo {title} {Topological axion states in the magnetic insulator
  ${\mathrm{mnbi}}_{2}{\mathrm{te}}_{4}$ with the quantized magnetoelectric
  effect},}\ }\href {\doibase 10.1103/PhysRevLett.122.206401} {\bibfield
  {journal} {\bibinfo  {journal} {Phys. Rev. Lett.}\ }\textbf {\bibinfo
  {volume} {122}},\ \bibinfo {pages} {206401} (\bibinfo {year}
  {2019})}\BibitemShut {NoStop}%
\bibitem [{\citenamefont {Li}\ \emph {et~al.}(2019)\citenamefont {Li},
  \citenamefont {Li}, \citenamefont {Du}, \citenamefont {Wang}, \citenamefont
  {Gu}, \citenamefont {Zhang}, \citenamefont {He}, \citenamefont {Duan},\ and\
  \citenamefont {Xu}}]{li2019}%
  \BibitemOpen
  \bibfield  {author} {\bibinfo {author} {\bibfnamefont {Jiaheng}\ \bibnamefont
  {Li}}, \bibinfo {author} {\bibfnamefont {Yang}\ \bibnamefont {Li}}, \bibinfo
  {author} {\bibfnamefont {Shiqiao}\ \bibnamefont {Du}}, \bibinfo {author}
  {\bibfnamefont {Zun}\ \bibnamefont {Wang}}, \bibinfo {author} {\bibfnamefont
  {Bing-Lin}\ \bibnamefont {Gu}}, \bibinfo {author} {\bibfnamefont
  {Shou-Cheng}\ \bibnamefont {Zhang}}, \bibinfo {author} {\bibfnamefont
  {Ke}~\bibnamefont {He}}, \bibinfo {author} {\bibfnamefont {Wenhui}\
  \bibnamefont {Duan}}, \ and\ \bibinfo {author} {\bibfnamefont {Yong}\
  \bibnamefont {Xu}},\ }\bibfield  {title} {\enquote {\bibinfo {title}
  {Intrinsic magnetic topological insulators in van der waals layered
  mnbi2te4-family materials},}\ }\href {\doibase 10.1126/sciadv.aaw5685}
  {\bibfield  {journal} {\bibinfo  {journal} {Sci. Adv.}\ }\textbf {\bibinfo
  {volume} {5}},\ \bibinfo {pages} {eaaw5685} (\bibinfo {year}
  {2019})}\BibitemShut {NoStop}%
\bibitem [{\citenamefont {Otrokov}\ \emph {et~al.}(2019)\citenamefont
  {Otrokov}, \citenamefont {Rusinov}, \citenamefont {Blanco-Rey}, \citenamefont
  {Hoffmann}, \citenamefont {Vyazovskaya}, \citenamefont {Eremeev},
  \citenamefont {Ernst}, \citenamefont {Echenique}, \citenamefont {Arnau},\
  and\ \citenamefont {Chulkov}}]{otrokov2019a}%
  \BibitemOpen
  \bibfield  {author} {\bibinfo {author} {\bibfnamefont {M.~M.}\ \bibnamefont
  {Otrokov}}, \bibinfo {author} {\bibfnamefont {I.~P.}\ \bibnamefont
  {Rusinov}}, \bibinfo {author} {\bibfnamefont {M.}~\bibnamefont {Blanco-Rey}},
  \bibinfo {author} {\bibfnamefont {M.}~\bibnamefont {Hoffmann}}, \bibinfo
  {author} {\bibfnamefont {A.~Yu.}\ \bibnamefont {Vyazovskaya}}, \bibinfo
  {author} {\bibfnamefont {S.~V.}\ \bibnamefont {Eremeev}}, \bibinfo {author}
  {\bibfnamefont {A.}~\bibnamefont {Ernst}}, \bibinfo {author} {\bibfnamefont
  {P.~M.}\ \bibnamefont {Echenique}}, \bibinfo {author} {\bibfnamefont
  {A.}~\bibnamefont {Arnau}}, \ and\ \bibinfo {author} {\bibfnamefont {E.~V.}\
  \bibnamefont {Chulkov}},\ }\bibfield  {title} {\enquote {\bibinfo {title}
  {Unique thickness-dependent properties of the van der waals interlayer
  antiferromagnet ${\mathrm{mnbi}}_{2}{\mathrm{te}}_{4}$ films},}\ }\href
  {\doibase 10.1103/PhysRevLett.122.107202} {\bibfield  {journal} {\bibinfo
  {journal} {Phys. Rev. Lett.}\ }\textbf {\bibinfo {volume} {122}},\ \bibinfo
  {pages} {107202} (\bibinfo {year} {2019})}\BibitemShut {NoStop}%
\bibitem [{\citenamefont {Chong}\ \emph {et~al.}(2020)\citenamefont {Chong},
  \citenamefont {Liu}, \citenamefont {Sharma}, \citenamefont {Kostin},
  \citenamefont {Gu}, \citenamefont {Fujita}, \citenamefont {Davis},\ and\
  \citenamefont {Sprau}}]{chong2020}%
  \BibitemOpen
  \bibfield  {author} {\bibinfo {author} {\bibfnamefont {Yi~Xue}\ \bibnamefont
  {Chong}}, \bibinfo {author} {\bibfnamefont {Xiaolong}\ \bibnamefont {Liu}},
  \bibinfo {author} {\bibfnamefont {Rahul}\ \bibnamefont {Sharma}}, \bibinfo
  {author} {\bibfnamefont {Andrey}\ \bibnamefont {Kostin}}, \bibinfo {author}
  {\bibfnamefont {Genda}\ \bibnamefont {Gu}}, \bibinfo {author} {\bibfnamefont
  {K.}~\bibnamefont {Fujita}}, \bibinfo {author} {\bibfnamefont
  {J.~C.~Séamus}\ \bibnamefont {Davis}}, \ and\ \bibinfo {author}
  {\bibfnamefont {Peter~O.}\ \bibnamefont {Sprau}},\ }\bibfield  {title}
  {\enquote {\bibinfo {title} {Severe dirac mass gap suppression in
  sb2te3-based quantum anomalous hall materials},}\ }\href {\doibase
  10.1021/acs.nanolett.0c02873} {\bibfield  {journal} {\bibinfo  {journal}
  {Nano Lett.}\ }\textbf {\bibinfo {volume} {20}},\ \bibinfo {pages}
  {8001--8007} (\bibinfo {year} {2020})}\BibitemShut {NoStop}%
\bibitem [{\citenamefont {Garnica}\ \emph {et~al.}(2022)\citenamefont
  {Garnica}, \citenamefont {Otrokov}, \citenamefont {Aguilar}, \citenamefont
  {Klimovskikh}, \citenamefont {Estyunin}, \citenamefont {Aliev}, \citenamefont
  {Amiraslanov}, \citenamefont {Abdullayev}, \citenamefont {Zverev},
  \citenamefont {Babanly}, \citenamefont {Mamedov}, \citenamefont {Shikin},
  \citenamefont {Arnau}, \citenamefont {de~Parga}, \citenamefont {Chulkov},\
  and\ \citenamefont {Miranda}}]{garnica2022}%
  \BibitemOpen
  \bibfield  {author} {\bibinfo {author} {\bibfnamefont {M.}~\bibnamefont
  {Garnica}}, \bibinfo {author} {\bibfnamefont {M.~M.}\ \bibnamefont
  {Otrokov}}, \bibinfo {author} {\bibfnamefont {P.~Casado}\ \bibnamefont
  {Aguilar}}, \bibinfo {author} {\bibfnamefont {I.~I.}\ \bibnamefont
  {Klimovskikh}}, \bibinfo {author} {\bibfnamefont {D.}~\bibnamefont
  {Estyunin}}, \bibinfo {author} {\bibfnamefont {Z.~S.}\ \bibnamefont {Aliev}},
  \bibinfo {author} {\bibfnamefont {I.~R.}\ \bibnamefont {Amiraslanov}},
  \bibinfo {author} {\bibfnamefont {N.~A.}\ \bibnamefont {Abdullayev}},
  \bibinfo {author} {\bibfnamefont {V.~N.}\ \bibnamefont {Zverev}}, \bibinfo
  {author} {\bibfnamefont {M.~B.}\ \bibnamefont {Babanly}}, \bibinfo {author}
  {\bibfnamefont {N.~T.}\ \bibnamefont {Mamedov}}, \bibinfo {author}
  {\bibfnamefont {A.~M.}\ \bibnamefont {Shikin}}, \bibinfo {author}
  {\bibfnamefont {A.}~\bibnamefont {Arnau}}, \bibinfo {author} {\bibfnamefont
  {A.~L.~Vázquez}\ \bibnamefont {de~Parga}}, \bibinfo {author} {\bibfnamefont
  {E.~V.}\ \bibnamefont {Chulkov}}, \ and\ \bibinfo {author} {\bibfnamefont
  {R.}~\bibnamefont {Miranda}},\ }\bibfield  {title} {\enquote {\bibinfo
  {title} {Native point defects and their implications for the dirac point gap
  at mnbi2te4(0001)},}\ }\href {\doibase 10.1038/s41535-021-00414-6} {\bibfield
   {journal} {\bibinfo  {journal} {npj Quantum Mater.}\ }\textbf {\bibinfo
  {volume} {7}},\ \bibinfo {pages} {7} (\bibinfo {year} {2022})}\BibitemShut
  {NoStop}%
\bibitem [{\citenamefont {Aroyo}\ \emph {et~al.}(2006)\citenamefont {Aroyo},
  \citenamefont {Perezmato}, \citenamefont {Capillas}, \citenamefont
  {Kroumova}, \citenamefont {Ivantchev}, \citenamefont {Madariaga},
  \citenamefont {Kirov},\ and\ \citenamefont {Wondratschek}}]{bilbao2}%
  \BibitemOpen
  \bibfield  {author} {\bibinfo {author} {\bibfnamefont {M~I}\ \bibnamefont
  {Aroyo}}, \bibinfo {author} {\bibfnamefont {J~M}\ \bibnamefont {Perezmato}},
  \bibinfo {author} {\bibfnamefont {C}~\bibnamefont {Capillas}}, \bibinfo
  {author} {\bibfnamefont {E}~\bibnamefont {Kroumova}}, \bibinfo {author}
  {\bibfnamefont {Svetoslav}\ \bibnamefont {Ivantchev}}, \bibinfo {author}
  {\bibfnamefont {G}~\bibnamefont {Madariaga}}, \bibinfo {author}
  {\bibfnamefont {A}~\bibnamefont {Kirov}}, \ and\ \bibinfo {author}
  {\bibfnamefont {Hans}\ \bibnamefont {Wondratschek}},\ }\bibfield  {title}
  {\enquote {\bibinfo {title} {Bilbao crystallographic server: I. databases and
  crystallographic computing programs},}\ }\href {\doibase
  10.1524/zkri.2006.221.1.15} {\bibfield  {journal} {\bibinfo  {journal} {Z.
  Krist.}\ }\textbf {\bibinfo {volume} {221}},\ \bibinfo {pages} {15--27}
  (\bibinfo {year} {2006})}\BibitemShut {NoStop}%
\bibitem [{\citenamefont {Kirov}\ \emph {et~al.}(2006)\citenamefont {Kirov},
  \citenamefont {Capillas}, \citenamefont {Perez-Mato},\ and\ \citenamefont
  {Wondratschek}}]{bilbao3}%
  \BibitemOpen
  \bibfield  {author} {\bibinfo {author} {\bibfnamefont {Asen}\ \bibnamefont
  {Kirov}}, \bibinfo {author} {\bibfnamefont {Cesar}\ \bibnamefont {Capillas}},
  \bibinfo {author} {\bibfnamefont {J}~\bibnamefont {Perez-Mato}}, \ and\
  \bibinfo {author} {\bibfnamefont {Hans}\ \bibnamefont {Wondratschek}},\
  }\bibfield  {title} {\enquote {\bibinfo {title} {Bilbao crystallographic
  server. ii. representations of crystallographic point groups and space
  groups},}\ }\href {\doibase 10.1107/S0108767305040286} {\bibfield  {journal}
  {\bibinfo  {journal} {Acta Cryst.}\ }\textbf {\bibinfo {volume} {62}},\
  \bibinfo {pages} {115--28} (\bibinfo {year} {2006})}\BibitemShut {NoStop}%
\bibitem [{\citenamefont {Perez-Mato}\ \emph {et~al.}(2011)\citenamefont
  {Perez-Mato}, \citenamefont {Orobengoa}, \citenamefont {Tasci}, \citenamefont
  {De~la Flor~Martin},\ and\ \citenamefont {Kirov}}]{bilbao1}%
  \BibitemOpen
  \bibfield  {author} {\bibinfo {author} {\bibfnamefont {J.}~\bibnamefont
  {Perez-Mato}}, \bibinfo {author} {\bibfnamefont {D}~\bibnamefont
  {Orobengoa}}, \bibinfo {author} {\bibfnamefont {Emre}\ \bibnamefont {Tasci}},
  \bibinfo {author} {\bibfnamefont {Gemma}\ \bibnamefont {De~la Flor~Martin}},
  \ and\ \bibinfo {author} {\bibfnamefont {A}~\bibnamefont {Kirov}},\
  }\bibfield  {title} {\enquote {\bibinfo {title} {Crystallography online:
  Bilbao crystallographic server},}\ }\href {\doibase
  https://www.researchgate.net/publication/228841764_Crystallography_Online_Bilbao_Crystallographic_Server}
  {\bibfield  {journal} {\bibinfo  {journal} {Bulg. Chem. Commun.}\ }\textbf
  {\bibinfo {volume} {43}},\ \bibinfo {pages} {183--197} (\bibinfo {year}
  {2011})}\BibitemShut {NoStop}%
\bibitem [{\citenamefont {Kruthoff}\ \emph {et~al.}(2017)\citenamefont
  {Kruthoff}, \citenamefont {de~Boer}, \citenamefont {van Wezel}, \citenamefont
  {Kane},\ and\ \citenamefont {Slager}}]{slager2017}%
  \BibitemOpen
  \bibfield  {author} {\bibinfo {author} {\bibfnamefont {Jorrit}\ \bibnamefont
  {Kruthoff}}, \bibinfo {author} {\bibfnamefont {Jan}\ \bibnamefont {de~Boer}},
  \bibinfo {author} {\bibfnamefont {Jasper}\ \bibnamefont {van Wezel}},
  \bibinfo {author} {\bibfnamefont {Charles~L.}\ \bibnamefont {Kane}}, \ and\
  \bibinfo {author} {\bibfnamefont {Robert-Jan}\ \bibnamefont {Slager}},\
  }\bibfield  {title} {\enquote {\bibinfo {title} {Topological classification
  of crystalline insulators through band structure combinatorics},}\ }\href
  {\doibase 10.1103/PhysRevX.7.041069} {\bibfield  {journal} {\bibinfo
  {journal} {Phys. Rev. X}\ }\textbf {\bibinfo {volume} {7}},\ \bibinfo {pages}
  {041069} (\bibinfo {year} {2017})}\BibitemShut {NoStop}%
\bibitem [{\citenamefont {Vergniory}\ \emph {et~al.}(2017)\citenamefont
  {Vergniory}, \citenamefont {Elcoro}, \citenamefont {Wang}, \citenamefont
  {Cano}, \citenamefont {Felser}, \citenamefont {Aroyo}, \citenamefont
  {Bernevig},\ and\ \citenamefont {Bradlyn}}]{vergniory2017}%
  \BibitemOpen
  \bibfield  {author} {\bibinfo {author} {\bibfnamefont {M.~G.}\ \bibnamefont
  {Vergniory}}, \bibinfo {author} {\bibfnamefont {L.}~\bibnamefont {Elcoro}},
  \bibinfo {author} {\bibfnamefont {Zhijun}\ \bibnamefont {Wang}}, \bibinfo
  {author} {\bibfnamefont {Jennifer}\ \bibnamefont {Cano}}, \bibinfo {author}
  {\bibfnamefont {C.}~\bibnamefont {Felser}}, \bibinfo {author} {\bibfnamefont
  {M.~I.}\ \bibnamefont {Aroyo}}, \bibinfo {author} {\bibfnamefont {B.~Andrei}\
  \bibnamefont {Bernevig}}, \ and\ \bibinfo {author} {\bibfnamefont {Barry}\
  \bibnamefont {Bradlyn}},\ }\bibfield  {title} {\enquote {\bibinfo {title}
  {Graph theory data for topological quantum chemistry},}\ }\href {\doibase
  10.1103/PhysRevE.96.023310} {\bibfield  {journal} {\bibinfo  {journal} {Phys.
  Rev. E}\ }\textbf {\bibinfo {volume} {96}},\ \bibinfo {pages} {023310}
  (\bibinfo {year} {2017})}\BibitemShut {NoStop}%
\bibitem [{\citenamefont {Elcoro}\ \emph {et~al.}(2017)\citenamefont {Elcoro},
  \citenamefont {Bradlyn}, \citenamefont {Wang}, \citenamefont {Vergniory},
  \citenamefont {Cano}, \citenamefont {Felser}, \citenamefont {Bernevig},
  \citenamefont {Orobengoa}, \citenamefont {Flor},\ and\ \citenamefont
  {Aroyo}}]{elcoro2017}%
  \BibitemOpen
  \bibfield  {author} {\bibinfo {author} {\bibfnamefont {L.}~\bibnamefont
  {Elcoro}}, \bibinfo {author} {\bibfnamefont {Barry}\ \bibnamefont {Bradlyn}},
  \bibinfo {author} {\bibfnamefont {Z.}~\bibnamefont {Wang}}, \bibinfo {author}
  {\bibfnamefont {M.~G.}\ \bibnamefont {Vergniory}}, \bibinfo {author}
  {\bibfnamefont {Jennifer}\ \bibnamefont {Cano}}, \bibinfo {author}
  {\bibfnamefont {C.}~\bibnamefont {Felser}}, \bibinfo {author} {\bibfnamefont
  {B.}~\bibnamefont {Bernevig}}, \bibinfo {author} {\bibfnamefont
  {D.}~\bibnamefont {Orobengoa}}, \bibinfo {author} {\bibfnamefont {G.~D.~L.}\
  \bibnamefont {Flor}}, \ and\ \bibinfo {author} {\bibfnamefont
  {M.}~\bibnamefont {Aroyo}},\ }\bibfield  {title} {\enquote {\bibinfo {title}
  {Double crystallographic groups and their representations on the bilbao
  crystallographic server},}\ }\href {\doibase 10.1107/S1600576717011712}
  {\bibfield  {journal} {\bibinfo  {journal} {J. Appl. Crystallogr}\ }\textbf
  {\bibinfo {volume} {50}},\ \bibinfo {pages} {1457} (\bibinfo {year}
  {2017})}\BibitemShut {NoStop}%
\bibitem [{\citenamefont {Bradlyn}\ \emph {et~al.}(2017)\citenamefont
  {Bradlyn}, \citenamefont {Elcoro}, \citenamefont {Cano}, \citenamefont
  {Vergniory}, \citenamefont {Wang}, \citenamefont {Felser}, \citenamefont
  {Aroyo},\ and\ \citenamefont {Bernevig}}]{bradlyn2017}%
  \BibitemOpen
  \bibfield  {author} {\bibinfo {author} {\bibfnamefont {Barry}\ \bibnamefont
  {Bradlyn}}, \bibinfo {author} {\bibfnamefont {L}~\bibnamefont {Elcoro}},
  \bibinfo {author} {\bibfnamefont {Jennifer}\ \bibnamefont {Cano}}, \bibinfo
  {author} {\bibfnamefont {MG}~\bibnamefont {Vergniory}}, \bibinfo {author}
  {\bibfnamefont {Zhijun}\ \bibnamefont {Wang}}, \bibinfo {author}
  {\bibfnamefont {C}~\bibnamefont {Felser}}, \bibinfo {author} {\bibfnamefont
  {MI}~\bibnamefont {Aroyo}}, \ and\ \bibinfo {author} {\bibfnamefont
  {B~Andrei}\ \bibnamefont {Bernevig}},\ }\bibfield  {title} {\enquote
  {\bibinfo {title} {Topological quantum chemistry},}\ }\href {\doibase
  10.1038/nature23268} {\bibfield  {journal} {\bibinfo  {journal} {Nature}\
  }\textbf {\bibinfo {volume} {547}},\ \bibinfo {pages} {298} (\bibinfo {year}
  {2017})}\BibitemShut {NoStop}%
\bibitem [{\citenamefont {Kresse}\ and\ \citenamefont
  {Furthm\"uller}(1996)}]{kresse1996}%
  \BibitemOpen
  \bibfield  {author} {\bibinfo {author} {\bibfnamefont {G.}~\bibnamefont
  {Kresse}}\ and\ \bibinfo {author} {\bibfnamefont {J.}~\bibnamefont
  {Furthm\"uller}},\ }\bibfield  {title} {\enquote {\bibinfo {title} {Efficient
  iterative schemes for ab initio total-energy calculations using a plane-wave
  basis set},}\ }\href {\doibase 10.1103/PhysRevB.54.11169} {\bibfield
  {journal} {\bibinfo  {journal} {Phys. Rev. B}\ }\textbf {\bibinfo {volume}
  {54}},\ \bibinfo {pages} {11169--11186} (\bibinfo {year} {1996})}\BibitemShut
  {NoStop}%
\bibitem [{\citenamefont {Perdew}\ \emph {et~al.}(1996)\citenamefont {Perdew},
  \citenamefont {Burke},\ and\ \citenamefont {Ernzerhof}}]{perdew1996}%
  \BibitemOpen
  \bibfield  {author} {\bibinfo {author} {\bibfnamefont {John~P.}\ \bibnamefont
  {Perdew}}, \bibinfo {author} {\bibfnamefont {Kieron}\ \bibnamefont {Burke}},
  \ and\ \bibinfo {author} {\bibfnamefont {Matthias}\ \bibnamefont
  {Ernzerhof}},\ }\bibfield  {title} {\enquote {\bibinfo {title} {Generalized
  gradient approximation made simple},}\ }\href {\doibase
  10.1103/PhysRevLett.77.3865} {\bibfield  {journal} {\bibinfo  {journal}
  {Phys. Rev. Lett.}\ }\textbf {\bibinfo {volume} {77}},\ \bibinfo {pages}
  {3865--3868} (\bibinfo {year} {1996})}\BibitemShut {NoStop}%
\bibitem [{\citenamefont {Dudarev}\ \emph {et~al.}(1998)\citenamefont
  {Dudarev}, \citenamefont {Botton}, \citenamefont {Savrasov}, \citenamefont
  {Humphreys},\ and\ \citenamefont {Sutton}}]{dudarev1998}%
  \BibitemOpen
  \bibfield  {author} {\bibinfo {author} {\bibfnamefont {S.~L.}\ \bibnamefont
  {Dudarev}}, \bibinfo {author} {\bibfnamefont {G.~A.}\ \bibnamefont {Botton}},
  \bibinfo {author} {\bibfnamefont {S.~Y.}\ \bibnamefont {Savrasov}}, \bibinfo
  {author} {\bibfnamefont {C.~J.}\ \bibnamefont {Humphreys}}, \ and\ \bibinfo
  {author} {\bibfnamefont {A.~P.}\ \bibnamefont {Sutton}},\ }\bibfield  {title}
  {\enquote {\bibinfo {title} {Electron-energy-loss spectra and the structural
  stability of nickel oxide: An lsda+u study},}\ }\href {\doibase
  10.1103/PhysRevB.57.1505} {\bibfield  {journal} {\bibinfo  {journal} {Phys.
  Rev. B}\ }\textbf {\bibinfo {volume} {57}},\ \bibinfo {pages} {1505--1509}
  (\bibinfo {year} {1998})}\BibitemShut {NoStop}%
\bibitem [{\citenamefont {Krukau}\ \emph {et~al.}(2006)\citenamefont {Krukau},
  \citenamefont {Vydrov}, \citenamefont {Izmaylov},\ and\ \citenamefont
  {Scuseria}}]{krukau2006}%
  \BibitemOpen
  \bibfield  {author} {\bibinfo {author} {\bibfnamefont {Aliaksandr~V.}\
  \bibnamefont {Krukau}}, \bibinfo {author} {\bibfnamefont {Oleg~A.}\
  \bibnamefont {Vydrov}}, \bibinfo {author} {\bibfnamefont {Artur~F.}\
  \bibnamefont {Izmaylov}}, \ and\ \bibinfo {author} {\bibfnamefont
  {Gustavo~E.}\ \bibnamefont {Scuseria}},\ }\bibfield  {title} {\enquote
  {\bibinfo {title} {Influence of the exchange screening parameter on the
  performance of screened hybrid functionals},}\ }\href {\doibase
  10.1063/1.2404663} {\bibfield  {journal} {\bibinfo  {journal} {J. Chem.
  Phys}\ }\textbf {\bibinfo {volume} {125}},\ \bibinfo {pages} {224106}
  (\bibinfo {year} {2006})}\BibitemShut {NoStop}%
\bibitem [{sup()}]{supple}%
  \BibitemOpen
  \href@noop {} {}\bibinfo {note} {See Supplementary Material at [url], for
  technical details on methods of first-principles calculation, stability and
  magnetic ground state, electronic structure and orbital projection,
  Ti$_2$Mo\textit{X}$_4$ and V$_2$Ta\textit{X}$_4$ family, bilayer and trilayer
  V$_2$WS$_4$, and theory, which includes
  Refs.~\cite{Blochl1994,grimme2010,mostofi2008,QuanSheng2018,gao2021,togo2015,Nose1984,Nose1991,Hoover1985,HE2021,Ozaki2003,Ozaki2004,Ozaki2005,Holstein1940,LiPing2019,Sui2020,Mellaerts2021}.}\BibitemShut
  {Stop}%
\bibitem [{\citenamefont {Crossland}\ \emph {et~al.}(2005)\citenamefont
  {Crossland}, \citenamefont {Hickey},\ and\ \citenamefont
  {Evans}}]{crossland2005}%
  \BibitemOpen
  \bibfield  {author} {\bibinfo {author} {\bibfnamefont {Clare~J.}\
  \bibnamefont {Crossland}}, \bibinfo {author} {\bibfnamefont {Peter~J.}\
  \bibnamefont {Hickey}}, \ and\ \bibinfo {author} {\bibfnamefont {John S.~O.}\
  \bibnamefont {Evans}},\ }\bibfield  {title} {\enquote {\bibinfo {title} {The
  synthesis and characterisation of cu2mx4 (m = w or mo; x = s{,} se or s/se)
  materials prepared by a solvothermal method},}\ }\href {\doibase
  10.1039/B507129A} {\bibfield  {journal} {\bibinfo  {journal} {J. Mater.
  Chem.}\ }\textbf {\bibinfo {volume} {15}},\ \bibinfo {pages} {3452--3458}
  (\bibinfo {year} {2005})}\BibitemShut {NoStop}%
\bibitem [{\citenamefont {Gan}\ and\ \citenamefont
  {Schwingenschl\"ogl}(2014)}]{gan2014}%
  \BibitemOpen
  \bibfield  {author} {\bibinfo {author} {\bibfnamefont {Li-Yong}\ \bibnamefont
  {Gan}}\ and\ \bibinfo {author} {\bibfnamefont {Udo}\ \bibnamefont
  {Schwingenschl\"ogl}},\ }\bibfield  {title} {\enquote {\bibinfo {title}
  {Two-dimensional square ternary cu${}_{2}$mx${}_{4}$ ($m$ = mo, w; $x$ = s,
  se) monolayers and nanoribbons predicted from density functional theory},}\
  }\href {\doibase 10.1103/PhysRevB.89.125423} {\bibfield  {journal} {\bibinfo
  {journal} {Phys. Rev. B}\ }\textbf {\bibinfo {volume} {89}},\ \bibinfo
  {pages} {125423} (\bibinfo {year} {2014})}\BibitemShut {NoStop}%
\bibitem [{\citenamefont {Hu}\ \emph {et~al.}(2016)\citenamefont {Hu},
  \citenamefont {Shao}, \citenamefont {Hang}, \citenamefont {Zhang},
  \citenamefont {Zhu},\ and\ \citenamefont {Xie}}]{hu2016}%
  \BibitemOpen
  \bibfield  {author} {\bibinfo {author} {\bibfnamefont {Xin}\ \bibnamefont
  {Hu}}, \bibinfo {author} {\bibfnamefont {Wei}\ \bibnamefont {Shao}}, \bibinfo
  {author} {\bibfnamefont {Xudong}\ \bibnamefont {Hang}}, \bibinfo {author}
  {\bibfnamefont {Xiaodong}\ \bibnamefont {Zhang}}, \bibinfo {author}
  {\bibfnamefont {Wenguang}\ \bibnamefont {Zhu}}, \ and\ \bibinfo {author}
  {\bibfnamefont {Yi}~\bibnamefont {Xie}},\ }\bibfield  {title} {\enquote
  {\bibinfo {title} {Superior electrical conductivity in hydrogenated layered
  ternary chalcogenide nanosheets for flexible all-solid-state
  supercapacitors},}\ }\href {\doibase https://doi.org/10.1002/anie.201600029}
  {\bibfield  {journal} {\bibinfo  {journal} {Angew. Chem. Int. Ed.}\ }\textbf
  {\bibinfo {volume} {55}},\ \bibinfo {pages} {5733--5738} (\bibinfo {year}
  {2016})}\BibitemShut {NoStop}%
\bibitem [{\citenamefont {Zhan}\ \emph {et~al.}(2018)\citenamefont {Zhan},
  \citenamefont {Wang}, \citenamefont {Li}, \citenamefont {Bo}, \citenamefont
  {Wang}, \citenamefont {Gao},\ and\ \citenamefont {Zhao}}]{zhan2018}%
  \BibitemOpen
  \bibfield  {author} {\bibinfo {author} {\bibfnamefont {Fengping}\
  \bibnamefont {Zhan}}, \bibinfo {author} {\bibfnamefont {Qinghua}\
  \bibnamefont {Wang}}, \bibinfo {author} {\bibfnamefont {Yibing}\ \bibnamefont
  {Li}}, \bibinfo {author} {\bibfnamefont {Xin}\ \bibnamefont {Bo}}, \bibinfo
  {author} {\bibfnamefont {Qingxiang}\ \bibnamefont {Wang}}, \bibinfo {author}
  {\bibfnamefont {Fei}\ \bibnamefont {Gao}}, \ and\ \bibinfo {author}
  {\bibfnamefont {Chuan}\ \bibnamefont {Zhao}},\ }\bibfield  {title} {\enquote
  {\bibinfo {title} {Low-temperature synthesis of cuboid silver
  tetrathiotungstate (ag2ws4) as electrocatalyst for hydrogen evolution
  reaction},}\ }\href {\doibase 10.1021/acs.inorgchem.8b00108} {\bibfield
  {journal} {\bibinfo  {journal} {Inorg. Chem.}\ }\textbf {\bibinfo {volume}
  {57}},\ \bibinfo {pages} {5791--5800} (\bibinfo {year} {2018})}\BibitemShut
  {NoStop}%
\bibitem [{\citenamefont {Wu}\ \emph {et~al.}(2019)\citenamefont {Wu},
  \citenamefont {Ma}, \citenamefont {Peng}, \citenamefont {Huang},\ and\
  \citenamefont {Dai}}]{wu2019}%
  \BibitemOpen
  \bibfield  {author} {\bibinfo {author} {\bibfnamefont {Qian}\ \bibnamefont
  {Wu}}, \bibinfo {author} {\bibfnamefont {Yandong}\ \bibnamefont {Ma}},
  \bibinfo {author} {\bibfnamefont {Rui}\ \bibnamefont {Peng}}, \bibinfo
  {author} {\bibfnamefont {Baibiao}\ \bibnamefont {Huang}}, \ and\ \bibinfo
  {author} {\bibfnamefont {Ying}\ \bibnamefont {Dai}},\ }\bibfield  {title}
  {\enquote {\bibinfo {title} {Single-layer cu2ws4 with promising
  electrocatalytic activity toward hydrogen evolution reaction},}\ }\href
  {\doibase 10.1021/acsami.9b18065} {\bibfield  {journal} {\bibinfo  {journal}
  {ACS Appl. Mater. Interfaces}\ }\textbf {\bibinfo {volume} {11}},\ \bibinfo
  {pages} {45818--45824} (\bibinfo {year} {2019})}\BibitemShut {NoStop}%
\bibitem [{\citenamefont {Lin}\ \emph {et~al.}(2019)\citenamefont {Lin},
  \citenamefont {Chen}, \citenamefont {Zhang},\ and\ \citenamefont
  {Song}}]{lin2019}%
  \BibitemOpen
  \bibfield  {author} {\bibinfo {author} {\bibfnamefont {Yunxiang}\
  \bibnamefont {Lin}}, \bibinfo {author} {\bibfnamefont {Shuangming}\
  \bibnamefont {Chen}}, \bibinfo {author} {\bibfnamefont {Ke}~\bibnamefont
  {Zhang}}, \ and\ \bibinfo {author} {\bibfnamefont {Li}~\bibnamefont {Song}},\
  }\bibfield  {title} {\enquote {\bibinfo {title} {Recent advance of ternary
  layered cu2mx4 (m=mo, w; x=s, se) nanomaterials for photocatalysis},}\ }\href
  {\doibase 10.1002/solr.201800320} {\bibfield  {journal} {\bibinfo  {journal}
  {Solar RRL}\ }\textbf {\bibinfo {volume} {3}},\ \bibinfo {pages} {1800320}
  (\bibinfo {year} {2019})}\BibitemShut {NoStop}%
\bibitem [{\citenamefont {Khomskii}(2004)}]{khomskii2004}%
  \BibitemOpen
  \bibfield  {author} {\bibinfo {author} {\bibfnamefont {Daniel~I.}\
  \bibnamefont {Khomskii}},\ }\href@noop {} {\emph {\bibinfo {title}
  {Transition Metal Compounds}}}\ (\bibinfo  {publisher} {Cambridge University
  Press},\ \bibinfo {year} {2004})\BibitemShut {NoStop}%
\bibitem [{\citenamefont {Wang}\ and\ \citenamefont {Wang}(2021)}]{wang2021}%
  \BibitemOpen
  \bibfield  {author} {\bibinfo {author} {\bibfnamefont {Huan}\ \bibnamefont
  {Wang}}\ and\ \bibinfo {author} {\bibfnamefont {Jing}\ \bibnamefont {Wang}},\
  }\bibfield  {title} {\enquote {\bibinfo {title} {Topological bands in
  two-dimensional orbital-active bipartite lattices},}\ }\href {\doibase
  10.1103/PhysRevB.103.L081109} {\bibfield  {journal} {\bibinfo  {journal}
  {Phys. Rev. B}\ }\textbf {\bibinfo {volume} {103}},\ \bibinfo {pages}
  {L081109} (\bibinfo {year} {2021})}\BibitemShut {NoStop}%
\bibitem [{\citenamefont {Fang}\ \emph {et~al.}(2012)\citenamefont {Fang},
  \citenamefont {Gilbert},\ and\ \citenamefont {Bernevig}}]{fang2012}%
  \BibitemOpen
  \bibfield  {author} {\bibinfo {author} {\bibfnamefont {Chen}\ \bibnamefont
  {Fang}}, \bibinfo {author} {\bibfnamefont {Matthew~J.}\ \bibnamefont
  {Gilbert}}, \ and\ \bibinfo {author} {\bibfnamefont {B.~Andrei}\ \bibnamefont
  {Bernevig}},\ }\bibfield  {title} {\enquote {\bibinfo {title} {Bulk
  topological invariants in noninteracting point group symmetric insulators},}\
  }\href {\doibase 10.1103/PhysRevB.86.115112} {\bibfield  {journal} {\bibinfo
  {journal} {Phys. Rev. B}\ }\textbf {\bibinfo {volume} {86}},\ \bibinfo
  {pages} {115112} (\bibinfo {year} {2012})}\BibitemShut {NoStop}%
\bibitem [{\citenamefont {Ikebe}\ \emph {et~al.}(2010)\citenamefont {Ikebe},
  \citenamefont {Morimoto}, \citenamefont {Masutomi}, \citenamefont {Okamoto},
  \citenamefont {Aoki},\ and\ \citenamefont {Shimano}}]{ikebe2010}%
  \BibitemOpen
  \bibfield  {author} {\bibinfo {author} {\bibfnamefont {Y.}~\bibnamefont
  {Ikebe}}, \bibinfo {author} {\bibfnamefont {T.}~\bibnamefont {Morimoto}},
  \bibinfo {author} {\bibfnamefont {R.}~\bibnamefont {Masutomi}}, \bibinfo
  {author} {\bibfnamefont {T.}~\bibnamefont {Okamoto}}, \bibinfo {author}
  {\bibfnamefont {H.}~\bibnamefont {Aoki}}, \ and\ \bibinfo {author}
  {\bibfnamefont {R.}~\bibnamefont {Shimano}},\ }\bibfield  {title} {\enquote
  {\bibinfo {title} {Optical hall effect in the integer quantum hall regime},}\
  }\href {\doibase 10.1103/PhysRevLett.104.256802} {\bibfield  {journal}
  {\bibinfo  {journal} {Phys. Rev. Lett.}\ }\textbf {\bibinfo {volume} {104}},\
  \bibinfo {pages} {256802} (\bibinfo {year} {2010})}\BibitemShut {NoStop}%
\bibitem [{\citenamefont {Shimano}\ \emph {et~al.}(2013)\citenamefont
  {Shimano}, \citenamefont {Yumoto}, \citenamefont {Yoo}, \citenamefont
  {Matsunaga}, \citenamefont {Tanabe}, \citenamefont {Hibino}, \citenamefont
  {Morimoto},\ and\ \citenamefont {Aoki}}]{shimano2013}%
  \BibitemOpen
  \bibfield  {author} {\bibinfo {author} {\bibfnamefont {R.}~\bibnamefont
  {Shimano}}, \bibinfo {author} {\bibfnamefont {G.}~\bibnamefont {Yumoto}},
  \bibinfo {author} {\bibfnamefont {J.~Y.}\ \bibnamefont {Yoo}}, \bibinfo
  {author} {\bibfnamefont {R.}~\bibnamefont {Matsunaga}}, \bibinfo {author}
  {\bibfnamefont {S.}~\bibnamefont {Tanabe}}, \bibinfo {author} {\bibfnamefont
  {H.}~\bibnamefont {Hibino}}, \bibinfo {author} {\bibfnamefont
  {T.}~\bibnamefont {Morimoto}}, \ and\ \bibinfo {author} {\bibfnamefont
  {H.}~\bibnamefont {Aoki}},\ }\bibfield  {title} {\enquote {\bibinfo {title}
  {Quantum faraday and kerr rotations in graphene},}\ }\href {\doibase
  10.1038/ncomms2866} {\bibfield  {journal} {\bibinfo  {journal} {Nature
  Commun.}\ }\textbf {\bibinfo {volume} {4}},\ \bibinfo {pages} {1841}
  (\bibinfo {year} {2013})}\BibitemShut {NoStop}%
\bibitem [{\citenamefont {Okada}\ \emph {et~al.}(2016)\citenamefont {Okada},
  \citenamefont {Takahashi}, \citenamefont {Mogi}, \citenamefont {Yoshimi},
  \citenamefont {Tsukazaki}, \citenamefont {Takahashi}, \citenamefont {Ogawa},
  \citenamefont {Kawasaki},\ and\ \citenamefont {Tokura}}]{okada2016}%
  \BibitemOpen
  \bibfield  {author} {\bibinfo {author} {\bibfnamefont {Ken~N.}\ \bibnamefont
  {Okada}}, \bibinfo {author} {\bibfnamefont {Youtarou}\ \bibnamefont
  {Takahashi}}, \bibinfo {author} {\bibfnamefont {Masataka}\ \bibnamefont
  {Mogi}}, \bibinfo {author} {\bibfnamefont {Ryutaro}\ \bibnamefont {Yoshimi}},
  \bibinfo {author} {\bibfnamefont {Atsushi}\ \bibnamefont {Tsukazaki}},
  \bibinfo {author} {\bibfnamefont {Kei~S.}\ \bibnamefont {Takahashi}},
  \bibinfo {author} {\bibfnamefont {Naoki}\ \bibnamefont {Ogawa}}, \bibinfo
  {author} {\bibfnamefont {Masashi}\ \bibnamefont {Kawasaki}}, \ and\ \bibinfo
  {author} {\bibfnamefont {Yoshinori}\ \bibnamefont {Tokura}},\ }\bibfield
  {title} {\enquote {\bibinfo {title} {Terahertz spectroscopy on faraday and
  kerr rotations in a quantum anomalous hall state},}\ }\href {\doibase
  10.1038/ncomms12245} {\bibfield  {journal} {\bibinfo  {journal} {Nature
  Commun.}\ }\textbf {\bibinfo {volume} {7}},\ \bibinfo {pages} {12245}
  (\bibinfo {year} {2016})}\BibitemShut {NoStop}%
\bibitem [{\citenamefont {Mogi}\ \emph {et~al.}(2022)\citenamefont {Mogi},
  \citenamefont {Okamura}, \citenamefont {Kawamura}, \citenamefont {Yoshimi},
  \citenamefont {Yasuda}, \citenamefont {Tsukazaki}, \citenamefont {Takahashi},
  \citenamefont {Morimoto}, \citenamefont {Nagaosa}, \citenamefont {Kawasaki},
  \citenamefont {Takahashi},\ and\ \citenamefont {Tokura}}]{mogi2022}%
  \BibitemOpen
  \bibfield  {author} {\bibinfo {author} {\bibfnamefont {M.}~\bibnamefont
  {Mogi}}, \bibinfo {author} {\bibfnamefont {Y.}~\bibnamefont {Okamura}},
  \bibinfo {author} {\bibfnamefont {M.}~\bibnamefont {Kawamura}}, \bibinfo
  {author} {\bibfnamefont {R.}~\bibnamefont {Yoshimi}}, \bibinfo {author}
  {\bibfnamefont {K.}~\bibnamefont {Yasuda}}, \bibinfo {author} {\bibfnamefont
  {A.}~\bibnamefont {Tsukazaki}}, \bibinfo {author} {\bibfnamefont {K.~S.}\
  \bibnamefont {Takahashi}}, \bibinfo {author} {\bibfnamefont {T.}~\bibnamefont
  {Morimoto}}, \bibinfo {author} {\bibfnamefont {N.}~\bibnamefont {Nagaosa}},
  \bibinfo {author} {\bibfnamefont {M.}~\bibnamefont {Kawasaki}}, \bibinfo
  {author} {\bibfnamefont {Y.}~\bibnamefont {Takahashi}}, \ and\ \bibinfo
  {author} {\bibfnamefont {Y.}~\bibnamefont {Tokura}},\ }\bibfield  {title}
  {\enquote {\bibinfo {title} {Experimental signature of the parity anomaly in
  a semi-magnetic topological insulator},}\ }\href {\doibase
  10.1038/s41567-021-01490-y} {\bibfield  {journal} {\bibinfo  {journal}
  {Nature Phys.}\ }\textbf {\bibinfo {volume} {18}},\ \bibinfo {pages}
  {390--394} (\bibinfo {year} {2022})}\BibitemShut {NoStop}%
\bibitem [{\citenamefont {Bl\"ochl}(1994)}]{Blochl1994}%
  \BibitemOpen
  \bibfield  {author} {\bibinfo {author} {\bibfnamefont {P.~E.}\ \bibnamefont
  {Bl\"ochl}},\ }\bibfield  {title} {\enquote {\bibinfo {title} {Projector
  augmented-wave method},}\ }\href {\doibase 10.1103/PhysRevB.50.17953}
  {\bibfield  {journal} {\bibinfo  {journal} {Phys. Rev. B}\ }\textbf {\bibinfo
  {volume} {50}},\ \bibinfo {pages} {17953--17979} (\bibinfo {year}
  {1994})}\BibitemShut {NoStop}%
\bibitem [{\citenamefont {Grimme}\ \emph {et~al.}(2010)\citenamefont {Grimme},
  \citenamefont {Antony}, \citenamefont {Ehrlich},\ and\ \citenamefont
  {Krieg}}]{grimme2010}%
  \BibitemOpen
  \bibfield  {author} {\bibinfo {author} {\bibfnamefont {Stefan}\ \bibnamefont
  {Grimme}}, \bibinfo {author} {\bibfnamefont {Jens}\ \bibnamefont {Antony}},
  \bibinfo {author} {\bibfnamefont {Stephan}\ \bibnamefont {Ehrlich}}, \ and\
  \bibinfo {author} {\bibfnamefont {Helge}\ \bibnamefont {Krieg}},\ }\bibfield
  {title} {\enquote {\bibinfo {title} {A consistent and accurate ab initio
  parametrization of density functional dispersion correction (dft-d) for the
  94 elements h-pu},}\ }\href {https://aip.scitation.org/doi/10.1063/1.3382344}
  {\bibfield  {journal} {\bibinfo  {journal} {J. Chem. Phys.}\ }\textbf
  {\bibinfo {volume} {132}},\ \bibinfo {pages} {154104} (\bibinfo {year}
  {2010})}\BibitemShut {NoStop}%
\bibitem [{\citenamefont {Mostofi}\ \emph {et~al.}(2008)\citenamefont
  {Mostofi}, \citenamefont {Yates}, \citenamefont {Lee}, \citenamefont {Souza},
  \citenamefont {Vanderbilt},\ and\ \citenamefont {Marzari}}]{mostofi2008}%
  \BibitemOpen
  \bibfield  {author} {\bibinfo {author} {\bibfnamefont {A.~A.}\ \bibnamefont
  {Mostofi}}, \bibinfo {author} {\bibfnamefont {J.~R.}\ \bibnamefont {Yates}},
  \bibinfo {author} {\bibfnamefont {Y.-S.}\ \bibnamefont {Lee}}, \bibinfo
  {author} {\bibfnamefont {I.}~\bibnamefont {Souza}}, \bibinfo {author}
  {\bibfnamefont {D.}~\bibnamefont {Vanderbilt}}, \ and\ \bibinfo {author}
  {\bibfnamefont {N.}~\bibnamefont {Marzari}},\ }\bibfield  {title} {\enquote
  {\bibinfo {title} {wannier90: A tool for obtaining maximally-localised
  wannier functions},}\ }\href {\doibase
  https://doi.org/10.1016/j.cpc.2007.11.016} {\bibfield  {journal} {\bibinfo
  {journal} {Comput. Phys. Commun.}\ }\textbf {\bibinfo {volume} {178}},\
  \bibinfo {pages} {685--699} (\bibinfo {year} {2008})}\BibitemShut {NoStop}%
\bibitem [{\citenamefont {Wu}\ \emph {et~al.}(2018)\citenamefont {Wu},
  \citenamefont {Zhang}, \citenamefont {Song}, \citenamefont {Troyer},\ and\
  \citenamefont {Soluyanov}}]{QuanSheng2018}%
  \BibitemOpen
  \bibfield  {author} {\bibinfo {author} {\bibfnamefont {Q.}~\bibnamefont
  {Wu}}, \bibinfo {author} {\bibfnamefont {S.}~\bibnamefont {Zhang}}, \bibinfo
  {author} {\bibfnamefont {H.-F.}\ \bibnamefont {Song}}, \bibinfo {author}
  {\bibfnamefont {M.}~\bibnamefont {Troyer}}, \ and\ \bibinfo {author}
  {\bibfnamefont {A.~A.}\ \bibnamefont {Soluyanov}},\ }\bibfield  {title}
  {\enquote {\bibinfo {title} {Wanniertools: An open-source software package
  for novel topological materials},}\ }\href {\doibase
  https://doi.org/10.1016/j.cpc.2017.09.033} {\bibfield  {journal} {\bibinfo
  {journal} {Comput. Phys. Commun.}\ }\textbf {\bibinfo {volume} {224}},\
  \bibinfo {pages} {405--416} (\bibinfo {year} {2018})}\BibitemShut {NoStop}%
\bibitem [{\citenamefont {Gao}\ \emph {et~al.}(2021)\citenamefont {Gao},
  \citenamefont {Wu}, \citenamefont {Persson},\ and\ \citenamefont
  {Wang}}]{gao2021}%
  \BibitemOpen
  \bibfield  {author} {\bibinfo {author} {\bibfnamefont {Jiacheng}\
  \bibnamefont {Gao}}, \bibinfo {author} {\bibfnamefont {Quansheng}\
  \bibnamefont {Wu}}, \bibinfo {author} {\bibfnamefont {Clas}\ \bibnamefont
  {Persson}}, \ and\ \bibinfo {author} {\bibfnamefont {Zhijun}\ \bibnamefont
  {Wang}},\ }\bibfield  {title} {\enquote {\bibinfo {title} {Irvsp: To obtain
  irreducible representations of electronic states in the vasp},}\ }\href
  {\doibase https://doi.org/10.1016/j.cpc.2020.107760} {\bibfield  {journal}
  {\bibinfo  {journal} {Comput. Phys. Commun.}\ }\textbf {\bibinfo {volume}
  {261}},\ \bibinfo {pages} {107760} (\bibinfo {year} {2021})}\BibitemShut
  {NoStop}%
\bibitem [{\citenamefont {Togo}\ and\ \citenamefont {Tanaka}(2015)}]{togo2015}%
  \BibitemOpen
  \bibfield  {author} {\bibinfo {author} {\bibfnamefont {Atsushi}\ \bibnamefont
  {Togo}}\ and\ \bibinfo {author} {\bibfnamefont {Isao}\ \bibnamefont
  {Tanaka}},\ }\bibfield  {title} {\enquote {\bibinfo {title} {First principles
  phonon calculations in materials science},}\ }\href
  {https://www.sciencedirect.com/science/article/pii/S1359646215003127}
  {\bibfield  {journal} {\bibinfo  {journal} {Scr. Mater.}\ }\textbf {\bibinfo
  {volume} {108}},\ \bibinfo {pages} {1--5} (\bibinfo {year}
  {2015})}\BibitemShut {NoStop}%
\bibitem [{\citenamefont {Nos{\'e}}(1984)}]{Nose1984}%
  \BibitemOpen
  \bibfield  {author} {\bibinfo {author} {\bibfnamefont {Shuichi}\ \bibnamefont
  {Nos{\'e}}},\ }\bibfield  {title} {\enquote {\bibinfo {title} {{A unified
  formulation of the constant temperature molecular dynamics methods}},}\
  }\href {\doibase 10.1063/1.447334} {\bibfield  {journal} {\bibinfo  {journal}
  {J. Chem. Phys.}\ }\textbf {\bibinfo {volume} {81}},\ \bibinfo {pages}
  {511--519} (\bibinfo {year} {1984})}\BibitemShut {NoStop}%
\bibitem [{\citenamefont {Nos{\'e}}(1991)}]{Nose1991}%
  \BibitemOpen
  \bibfield  {author} {\bibinfo {author} {\bibfnamefont {Shuichi}\ \bibnamefont
  {Nos{\'e}}},\ }\bibfield  {title} {\enquote {\bibinfo {title} {{Constant
  Temperature Molecular Dynamics Methods}},}\ }\href {\doibase
  10.1143/PTPS.103.1} {\bibfield  {journal} {\bibinfo  {journal} {Prog. Theor.
  Phys. Supp.}\ }\textbf {\bibinfo {volume} {103}},\ \bibinfo {pages} {1--46}
  (\bibinfo {year} {1991})}\BibitemShut {NoStop}%
\bibitem [{\citenamefont {Hoover}(1985)}]{Hoover1985}%
  \BibitemOpen
  \bibfield  {author} {\bibinfo {author} {\bibfnamefont {William~G.}\
  \bibnamefont {Hoover}},\ }\bibfield  {title} {\enquote {\bibinfo {title}
  {Canonical dynamics: Equilibrium phase-space distributions},}\ }\href
  {\doibase 10.1103/PhysRevA.31.1695} {\bibfield  {journal} {\bibinfo
  {journal} {Phys. Rev. A}\ }\textbf {\bibinfo {volume} {31}},\ \bibinfo
  {pages} {1695--1697} (\bibinfo {year} {1985})}\BibitemShut {NoStop}%
\bibitem [{\citenamefont {He}\ \emph {et~al.}(2021)\citenamefont {He},
  \citenamefont {Helbig}, \citenamefont {Verstraete},\ and\ \citenamefont
  {Bousquet}}]{HE2021}%
  \BibitemOpen
  \bibfield  {author} {\bibinfo {author} {\bibfnamefont {Xu}~\bibnamefont
  {He}}, \bibinfo {author} {\bibfnamefont {Nicole}\ \bibnamefont {Helbig}},
  \bibinfo {author} {\bibfnamefont {Matthieu~J.}\ \bibnamefont {Verstraete}}, \
  and\ \bibinfo {author} {\bibfnamefont {Eric}\ \bibnamefont {Bousquet}},\
  }\bibfield  {title} {\enquote {\bibinfo {title} {Tb2j: A python package for
  computing magnetic interaction parameters},}\ }\href {\doibase
  https://doi.org/10.1016/j.cpc.2021.107938} {\bibfield  {journal} {\bibinfo
  {journal} {Comput. Phys. Commun.}\ }\textbf {\bibinfo {volume} {264}},\
  \bibinfo {pages} {107938} (\bibinfo {year} {2021})}\BibitemShut {NoStop}%
\bibitem [{\citenamefont {Ozaki}(2003)}]{Ozaki2003}%
  \BibitemOpen
  \bibfield  {author} {\bibinfo {author} {\bibfnamefont {T.}~\bibnamefont
  {Ozaki}},\ }\bibfield  {title} {\enquote {\bibinfo {title} {Variationally
  optimized atomic orbitals for large-scale electronic structures},}\ }\href
  {\doibase 10.1103/PhysRevB.67.155108} {\bibfield  {journal} {\bibinfo
  {journal} {Phys. Rev. B}\ }\textbf {\bibinfo {volume} {67}},\ \bibinfo
  {pages} {155108} (\bibinfo {year} {2003})}\BibitemShut {NoStop}%
\bibitem [{\citenamefont {Ozaki}\ and\ \citenamefont {Kino}(2004)}]{Ozaki2004}%
  \BibitemOpen
  \bibfield  {author} {\bibinfo {author} {\bibfnamefont {T.}~\bibnamefont
  {Ozaki}}\ and\ \bibinfo {author} {\bibfnamefont {H.}~\bibnamefont {Kino}},\
  }\bibfield  {title} {\enquote {\bibinfo {title} {Numerical atomic basis
  orbitals from h to kr},}\ }\href {\doibase 10.1103/PhysRevB.69.195113}
  {\bibfield  {journal} {\bibinfo  {journal} {Phys. Rev. B}\ }\textbf {\bibinfo
  {volume} {69}},\ \bibinfo {pages} {195113} (\bibinfo {year}
  {2004})}\BibitemShut {NoStop}%
\bibitem [{\citenamefont {Ozaki}\ and\ \citenamefont {Kino}(2005)}]{Ozaki2005}%
  \BibitemOpen
  \bibfield  {author} {\bibinfo {author} {\bibfnamefont {T.}~\bibnamefont
  {Ozaki}}\ and\ \bibinfo {author} {\bibfnamefont {H.}~\bibnamefont {Kino}},\
  }\bibfield  {title} {\enquote {\bibinfo {title} {Efficient projector
  expansion for the ab initio lcao method},}\ }\href {\doibase
  10.1103/PhysRevB.72.045121} {\bibfield  {journal} {\bibinfo  {journal} {Phys.
  Rev. B}\ }\textbf {\bibinfo {volume} {72}},\ \bibinfo {pages} {045121}
  (\bibinfo {year} {2005})}\BibitemShut {NoStop}%
\bibitem [{\citenamefont {Holstein}\ and\ \citenamefont
  {Primakoff}(1940)}]{Holstein1940}%
  \BibitemOpen
  \bibfield  {author} {\bibinfo {author} {\bibfnamefont {T.}~\bibnamefont
  {Holstein}}\ and\ \bibinfo {author} {\bibfnamefont {H.}~\bibnamefont
  {Primakoff}},\ }\bibfield  {title} {\enquote {\bibinfo {title} {Field
  dependence of the intrinsic domain magnetization of a ferromagnet},}\ }\href
  {\doibase 10.1103/PhysRev.58.1098} {\bibfield  {journal} {\bibinfo  {journal}
  {Phys. Rev.}\ }\textbf {\bibinfo {volume} {58}},\ \bibinfo {pages}
  {1098--1113} (\bibinfo {year} {1940})}\BibitemShut {NoStop}%
\bibitem [{\citenamefont {Li}(2019)}]{LiPing2019}%
  \BibitemOpen
  \bibfield  {author} {\bibinfo {author} {\bibfnamefont {Ping}\ \bibnamefont
  {Li}},\ }\bibfield  {title} {\enquote {\bibinfo {title} {Prediction of
  intrinsic two dimensional ferromagnetism realized quantum anomalous hall
  effect},}\ }\href {\doibase 10.1039/C8CP07781A} {\bibfield  {journal}
  {\bibinfo  {journal} {Phys. Chem. Chem. Phys.}\ }\textbf {\bibinfo {volume}
  {21}},\ \bibinfo {pages} {6712--6717} (\bibinfo {year} {2019})}\BibitemShut
  {NoStop}%
\bibitem [{\citenamefont {Sui}\ \emph {et~al.}(2020)\citenamefont {Sui},
  \citenamefont {Zhang}, \citenamefont {Jin}, \citenamefont {Xia},\ and\
  \citenamefont {Li}}]{Sui2020}%
  \BibitemOpen
  \bibfield  {author} {\bibinfo {author} {\bibfnamefont {Qian}\ \bibnamefont
  {Sui}}, \bibinfo {author} {\bibfnamefont {Jiaxin}\ \bibnamefont {Zhang}},
  \bibinfo {author} {\bibfnamefont {Suhua}\ \bibnamefont {Jin}}, \bibinfo
  {author} {\bibfnamefont {Yunyouyou}\ \bibnamefont {Xia}}, \ and\ \bibinfo
  {author} {\bibfnamefont {Gang}\ \bibnamefont {Li}},\ }\bibfield  {title}
  {\enquote {\bibinfo {title} {Model hamiltonian for the quantum anomalous hall
  state in iron-halogenide},}\ }\href {\doibase 10.1088/0256-307X/37/9/097301}
  {\bibfield  {journal} {\bibinfo  {journal} {Chin. Phys. Lett.}\ }\textbf
  {\bibinfo {volume} {37}},\ \bibinfo {pages} {097301} (\bibinfo {year}
  {2020})}\BibitemShut {NoStop}%
\bibitem [{\citenamefont {Mellaerts}\ \emph {et~al.}(2021)\citenamefont
  {Mellaerts}, \citenamefont {Meng}, \citenamefont {Afanasiev}, \citenamefont
  {Seo}, \citenamefont {Houssa},\ and\ \citenamefont
  {Locquet}}]{Mellaerts2021}%
  \BibitemOpen
  \bibfield  {author} {\bibinfo {author} {\bibfnamefont {S.}~\bibnamefont
  {Mellaerts}}, \bibinfo {author} {\bibfnamefont {R.}~\bibnamefont {Meng}},
  \bibinfo {author} {\bibfnamefont {V.}~\bibnamefont {Afanasiev}}, \bibinfo
  {author} {\bibfnamefont {J.~W.}\ \bibnamefont {Seo}}, \bibinfo {author}
  {\bibfnamefont {M.}~\bibnamefont {Houssa}}, \ and\ \bibinfo {author}
  {\bibfnamefont {J.-P.}\ \bibnamefont {Locquet}},\ }\bibfield  {title}
  {\enquote {\bibinfo {title} {Quarter-filled kane-mele hubbard model: Dirac
  half metals},}\ }\href {\doibase 10.1103/PhysRevB.103.155159} {\bibfield
  {journal} {\bibinfo  {journal} {Phys. Rev. B}\ }\textbf {\bibinfo {volume}
  {103}},\ \bibinfo {pages} {155159} (\bibinfo {year} {2021})}\BibitemShut
  {NoStop}%
\end{thebibliography}
\end{document}